\documentclass[letterpaper,twocolumn,10pt]{article}
\def\arxiv{a}
\def\camera{c}
\def\mode{a}
\usepackage{usenix2019_v3}
\usepackage{authblk}
\usepackage{outlines}
\usepackage{xspace}
\usepackage{tikz}
\usepackage{amsmath,epsfig}
\usepackage{caption, subcaption}
\usepackage{algpseudocode}
\usepackage{algorithm}
\usepackage{cleveref}
\usepackage{enumitem}
\usepackage{url}
\usepackage{hyperref}
\usepackage[normalem]{ulem} 

\usepackage{amssymb,amsmath, mathtools}

\usepackage{filecontents} 
\usepackage{multirow}
\usepackage{comment}

\usepackage{footnote}
\usepackage{footnotebackref}
\usepackage{adjustbox}

\Crefname{figure}{Fig.}{Figs.}
\setlist[itemize]{leftmargin=*,noitemsep,nolistsep}

\newcommand{\cc}{CC\xspace} 
\newcommand{\sys}{Polyjuice\xspace}

\newcommand{\RW}{$\xrightarrow{rw}$\xspace}
\newcommand{\WR}{$\xrightarrow{wr}$\xspace}
\newcommand{\WW}{$\xrightarrow{ww}$\xspace}

\newcommand{\changeremove}[1]{}

\begin{document}

\date{}

\title{\Large \bf \sys: High-Performance Transactions via Learned Concurrency Control}

\author[$^{\dagger\ast}$]{Jiachen Wang}
\author[$\ddagger\ast$]{Ding Ding}
\author[$^{\dagger}$]{Huan Wang}
\author[$\ddagger$]{Conrad Christensen}
\author[$^{\dagger}$]{Zhaoguo Wang}
\author[$^{\dagger}$]{Haibo Chen}
\author[$\ddagger$]{Jinyang Li}

\affil[$\dagger$]{%
  Institute of Parallel and Distributed Systems, Shanghai Jiao Tong University
}

\affil[$\dagger$]{%
  Shanghai AI Laboratory
}

\affil[$\dagger$]{%
  Engineering Research Center for Domain-specific Operating Systems, Ministry of Education, China
}

\affil[$\ddagger$]{%
  Department of Computer Science, New York University
}

\pagestyle{empty}  
\thispagestyle{empty}
\maketitle

\renewcommand{\thefootnote}{}
\begin{abstract}
    Concurrency control algorithms are key determinants of the performance of in-memory databases. Existing algorithms are designed to work well for certain workloads. For example, optimistic concurrency control (OCC) is better than two-phase-locking (2PL) under low contention, while the converse is true under high contention.
    
    To adapt to different workloads, prior works mix or switch between a few known algorithms using manual insights or simple heuristics.  We propose a learning-based framework that instead explicitly optimizes concurrency control via offline training to maximize performance. Instead of choosing among a small number of known algorithms, our approach searches in a ``policy space'' of fine-grained actions, resulting in novel algorithms that can outperform existing algorithms by specializing to a given workload.
    
    We build \sys \footnote{$\ast$ Jiachen Wang and Ding Ding contributed equally to this paper.} based on our learning framework and evaluate it against several existing algorithms. Under different configurations of TPC-C and TPC-E, \sys can achieve throughput numbers higher than the best of existing algorithms by 15\% to 56\%.
\end{abstract}

\section{Introduction}
Concurrency control (\cc) algorithms lie at the foundation of modern database systems~\cite{gray:book}. 
A \cc algorithm synchronizes a
transaction's access to storage objects to maximize concurrent execution while guaranteeing 
correctness.  As today's database systems are 
no longer disk-bound, the \cc algorithm in use becomes crucial to 
a database's performance.

\begin{figure}[t]
\centering
\includegraphics[scale = 0.8]{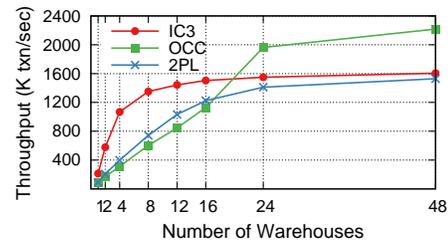}
\caption{IC3, OCC, 2PL performance on TPC-C, 48 threads.}
\label{fig:intro-tpcc-performance}
\end{figure}

Traditional \cc algorithms, such as two-phase-locking (2PL)~\cite{2pl:gray75} and optimistic concurrency control (OCC)~\cite{occ:kung81}, take fixed algorithmic steps regardless of the workload. Thus, it comes as no surprise that the relative performance of different algorithms varies depending on  
the transaction workload.  
Figure~\ref{fig:intro-tpcc-performance} shows the throughput of 2PL, OCC and IC3~\cite{wang2016scaling} on a multi-core database under the TPC-C workload with a varying number of warehouses. OCC has the highest throughput under low contention (more warehouses) while the other two outperform OCC under high contention (fewer warehouses). Similar results have also been reported by others~\cite{abyss:vldb14}. 

To adapt to different workloads, prior works 
propose a federated approach by simultaneously supporting a small number of existing \cc algorithms, including 2PL and OCC.  These systems require users to partition the workload either by data~\cite{tang2018toward} or by transaction type~\cite{su2017bringing, xie2015high, modularcc:sha88}. The decision of which algorithm to use for each partition is either based on manual insights~\cite{su2017bringing, xie2015high, modularcc:sha88} or simple runtime metrics~\cite{tang2018toward}.  While this federated approach can improve performance, it has limitations. 
First, by limiting itself to using a small number of known algorithms, it lacks the flexibility to customize concurrency control to fully exploit the workload.  Second, by relying on manual insights or simple heuristics, it lacks a systematic solution to optimize concurrency control for performance.

This paper presents a learning-based framework to optimize concurrency control for a given workload. We assume that the workload is known a priori such as past workloads, e.g. 
in the form of stored procedures. To enable learning, we design a ``policy space'' of fine-grained actions (a.k.a. algorithmic steps): each policy can be viewed as a \cc algorithm that uses specific actions to synchronize different data accesses 
made by different transactions.  All policies perform an explicit validation before transactions commit to ensure serializability. We use offline training to learn the highest performing policy for a given workload.  This framework is expressive: it can learn new \cc algorithms as well as existing ones. It also allows explicit optimization for performance via systematic searches of the policy space. 

We have realized the design for learned concurrency control in a system called \sys for multi-core in-memory databases.
The core technical challenge of \sys is to design the policy space.  Inspired by reinforcement learning, we view each policy as a function that maps each state (i.e., the execution context of a data access) to actions that control the interleavings of accesses made by concurrent transactions. 
In \sys, the state specifies what type of transaction is being used and which of its accesses are under execution.  The actions support multiple ways of interleaving control, including deciding which data version to read, whether to expose an uncommitted write, how long to wait before access, and whether to perform early validation before commit.

\sys represents each policy using a table: the rows correspond to different states and the columns correspond to different kinds of actions.  
\sys uses evolutionary algorithm based training to search the policy space for the policy that has the highest commit throughput for a given workload. 

We train and evaluate \sys's performance on micro-benchmarks, TPC-C and TPC-E, and compare with existing algorithms, including Silo~\cite{silo:sosp13}(OCC), 2PL, Tebaldi~\cite{su2017bringing}, CormCC~\cite{tang2018toward} and IC3~\cite{wang2016scaling}.  Our experiments show that, for TPC-C and TPC-E with moderate to high contention, \sys can find a \cc policy whose throughput is better than the best of existing algorithms by $15\%$ to $56\%$. Detailed analysis shows that \sys can learn an interesting policy that is different than any of the existing algorithms to exploit the workload in subtle ways(\S\ref{subsec:case-study}).  For workloads with almost no contention, \sys learns the same policy as OCC and incurs $8\%$ slowdown due to its implementation overhead. 

As \sys requires offline training, it is not suitable for dynamic workloads that can change rapidly and unpredictably.  However, our analysis of an e-commerce website trace shows that real-world workloads are fairly predictable in terms of its peak hour workload characteristics including the likelihood of conflict.  This suggests that it is practical to use \sys to optimize a database's peak performance by training on traces of recently observed peak workloads.

In summary, our paper makes the follow contributions:
\begin{itemize}
    \item We present the first framework to
    learn concurrency control using a policy space of fine-grained actions.
    \item We design \sys's policy space according to the framework so that it can encode a variety of existing \cc algorithms while allowing the exploration of new ones.
    \item We show that \sys's policy, represented as a table, can be optimized simply using an evolutionary algorithm. 
    \item Even for the heavily-studied TPC-C benchmarks, \sys can find interesting and novel policies not seen in existing algorithms to improve transaction throughput under moderate to high contention.
\end{itemize}

\section{Background and Motivation}
\label{sec:background}

Existing works have realized the inadequacy of using one fixed concurrency control algorithm for different workloads.
For the solution, they propose a federated approach of mixing a few (typically 2 or 3) known \cc algorithms~\cite{tang2018toward,xie2015high, su2017bringing, wang2016mostly}.  In this section, we discuss the limitations of this federated approach and motivate the need for a more expressive learning-based approach.

The federated approach of adapting \cc to a workload is characterized by its {\em coarse-grained} way of mixing different algorithms. Specifically, this approach coarsely partitions the workload.  The same \cc algorithm is used within a workload partition, while a different algorithm may be used for a different partition. Two ways of partitioning can be found in existing work.  CormCC~\cite{tang2018toward} partitions by data: all accesses to data in the same partition use the same \cc algorithm.  Tebaldi~\cite{su2017bringing} and Callas~\cite{xie2015high} group (a.k.a. partition) transactions by types: all transactions belonging to the same group (a.k.a. partition) use the same \cc for all their data accesses. 

The coarse-grained way of mixing \cc algorithms is limited in its ability to fully exploit workload characteristics for performance.  For example, with CormCC, if transactions $T$ and $T'$ both only access data within the same partition, they would synchronize all of their accesses using the same \cc algorithm. Similarly for Tebaldi and Callas, if transactions $T$ and $T'$ are of the same type, they would always use the same \cc algorithm. This is not optimal: if different data accesses of $T$ and $T'$ have different contention characteristics, they may be better served by different methods for controlling concurrency. 

A second limitation of existing federated \cc work is their reliance on manual insights to partition the workload or to determine which \cc algorithm to use for each partition.
Callas and Tebaldi manually assign transactions to groups and choose a specific \cc algorithm for each group. CormCC partitions the TPC-C workload by warehouse based on manual insights and uses simple runtime statistics (e.g. read/write ratio) to decide which \cc algorithm to use for each partition.

\vspace{0.1in}
\noindent{\bf Our approach.} We aim to optimize \cc for a given workload in a {\em fine-grained} way using a learning-based approach.  Instead of partitioning the workload and using a single \cc algorithm for all data accesses within the partition, we propose to allow each data access to use one of many different fine-grained ``actions'' to mediate potentially conflicting accesses.  
When deciding what action(s) to take to maximize performance, we are not concerned with correctness; instead, we rely on a separate validation mechanism to abort non-serializable transactions.
As fine-grained actions lead to exponentially many choices for a given workload, it is impossible to rely on manual insights to choose the best action(s). A more practical solution is to use a learning-based approach to explicitly optimize the choice of actions for the given workload.

The main challenge of our approach is to design the learning framework with fine-grained actions for concurrency control. Ideally, the framework should be expressive enough to encode {\bf most} existing \cc algorithms and to allow the synthesis of new ones.  In the next section, we discuss how to design such a learning framework.

\section{Learning Concurrency Control}
\label{sec:motivation}

In this section, we examine how to frame concurrency control as a fine-grained learning task.

\vspace{0.1in}
\noindent \textbf{System settings.} Our target setting is an in-memory database running on a single multi-core machine.  We assume the kinds of transaction to be run on the database are known a priori, e.g. in the form of stored procedures. 
A number of existing work also exploit a known-workload in designing \cc algorithms~\cite{xie2015high, wang2016scaling, mu2019deferred}.
Our work focuses on learning concurrency control for read-write transactions, and reuses existing mechanisms to support logging and 
snapshot-based read-only transactions~\cite{silo:sosp13}. Although our learning framework is general enough to represent multi-version concurrency control (MVCC), our later system design does not support it because existing snapshot-based read-only transactions can already capture much of MVCC's performance benefits.

\subsection{The learning framework} Our framework for learning concurrency control is inspired by reinforcement learning (RL). As one of the major branches of machine learning, RL involves learning how to interact with an environment to maximize a numerical reward. The key ingredients in RL are: a {\em policy} that maps perceived states of the environment to actions to be taken when those states are reached, a {\em reward} signal that defines the optimization goal, and the {\em environment} under which the learning system operates. In our context, the policy corresponds to the \cc algorithm; the reward corresponds to some performance metric to be maximized; the environment captures the 
transaction workload and system setup under which the \cc operates.  

It is straightforward to decide on the optimization objective (a.k.a. reward). In this work, we use transaction throughput. Compared to latency or abort rate, transaction throughput is widely used as the key end-to-end performance metric for in-memory databases.

It is non-trivial to design a ``policy space'' to represent various \cc algorithms.  
At a high level, a \cc algorithm executes a transaction by controlling how its data access can interleave with potentially conflicting accesses from other concurrent transactions.  As mentioned previously, we do not attempt to learn how to guarantee correctness.  Instead, a learned \cc algorithm always invokes a manually-designed validation procedure as part of transaction commit to ensure serializability. What we do learn is a policy that determines what actions to take in order to maximize performance for a given workload.   A good \cc policy balances how long transactions execute vs. how likely transactions are aborted, resulting in 
a high reward, as measured by how many transactions successfully commit per second.  Aside from the \cc policy, how long a database backs off before retrying an aborted transaction can also affect the performance. We separate the backoff policy from the \cc policy, and this section focuses on the latter. 

\vspace{0.1in}
\noindent{\bf The policy space of concurrency control.} 
Taking a page from reinforcement learning, we represent the policy as a mapping from some state of execution to a specific action to take upon encountering that state. Taking different actions in different states 
allows us to specialize a \cc algorithm to optimize for a given workload. Thus, the state space should include information that is necessary to distinguish circumstances 
that require different actions, e.g. the type of transaction that is making the access, the type of access etc. In a later section (\S\ref{subsec:state}), we provide a concrete design of the state space.  In the rest of this section, we focus on designing the action space.

\begin{table*}[h]
  \small
  \centering
  \begin{tabular}{r||c|c|c|c||c|c}
    & \multicolumn{4}{c||}{\bf{Interleaving control}} & \multicolumn{2}{c}{\bf{Validation}} \\ 
    & Read & Read & Write & Write  &
    Early  & Validation \\ 
    & wait & version & wait & visibility & validation & method\\\hline \hline
    
    \multirow{2}{*}{2PL$^*$} & Until $T_{dep}$ & latest & Until $T_{dep}$ & \multirow{2}{*}{Yes} & \multirow{2}{*}{Yes} & \multirow{2}{*}{n/a} \\ 
    & commits & committed & commits & & & \\\hline
    
    OCC~\cite{occ:kung81} & \multirow{2}{*}{No} & latest & \multirow{2}{*}{No} & \multirow{2}{*}{No} & No & physical cts\\ 
    TicToc~\cite{yu2016tictoc} & & committed & & & No & logical cts \\ \hline
    
    \multirow{2}{*}{Sundial~\cite{sundial:vldb18}} & \multirow{2}{*}{No} & latest & Until $T_{wdep}$ & \multirow{2}{*}{No} & \multirow{2}{*}{No} & \multirow{2}{*}{logical cts} \\
    & & committed & commits & & & \\\hline
    
    Callas RP~\cite{xie2015high} & Until $T_{dep}$ finish &  latest & Until $T_{dep}$ finish & \multirow{2}{*}{piece-end}
    & \multirow{2}{*}{piece-end}
    & n/a or\\
    
    IC3~\cite{wang2016scaling}, DRP~\cite{mu2019deferred} & 
    certain access & un-committed & certain access & & & physical cts \\ \hline
    
     
     MVTSO~\cite{database:bernstein} & Until $T'\in T_{wdep}$ commits &  largest committed & \multirow{2}{*}{No} & \multirow{2}{*}{Yes} & \multirow{2}{*}{Yes} & \multirow{2}{*}{physical ts} \\ 
     
     (MVCC) & if $ts(T')<ts(T)$ & $< ts(T)$ & & & & \\
  \end{tabular}
  
  
  \caption{The choices made in existing \cc algorithms according to the action space described in \S\ref{sec:motivation}. $T$ refers to the current transaction. $T_{dep}$ refers to the set of transactions that $T$ is dependent on (due to its conflicting access so far). $T_{wdep}$ is the subset of $T_{dep}$ whose writes have conflicted with $T$.  $ts(T)$ refers to the timestamp assigned to $T$ by MVTSO~\cite{database:bernstein}.}
  \label{table:classification}
\end{table*}

Ideally, the action space should encompass a set of fine-grained actions that can be mixed and matched to represent many different \cc algorithms. 
These actions can be classified into two categories:
1) actions that control how the data access of concurrent transactions can interleave during transaction execution, and 2) actions that control when and how to perform validation in order to detect whether an executed transaction has violated serializability. 
Next, we discuss the spectrum of actions available to use in each of the two categories.

\vspace{0.1in}
\noindent \textbf{Available actions for interleaving control.} These actions mediate potentially conflicting data accesses, thereby affecting the set of dependencies that arise among concurrent transactions.   There are 3 types of dependencies: write-write \WW (a.k.a. write dependency), write-read \WR 
(a.k.a. read dependency), or read-write \RW (a.k.a. anti-dependency)~\cite{atul:thesis}. What are the knobs of control that can affect these dependencies?

To discover these knobs in their full generality, let us assume a hypothetical yet still practical database design that keeps track of each read and write access of transactions in a per-object access list, similar to the approach taken in~\cite{mu2014extracting,wang2016scaling}. As a transaction $T$ performs data accesses, it may insert its reads/writes to the corresponding per-object access lists while also updating $T_{dep}$, the set of transactions that $T$ becomes dependent on. Using this flexible way of tracking dependencies enables a wide range of design choices for interleaving control, as we will see next.

When executing transaction $T$, a \cc algorithm has the following action choices:
\begin{itemize}
\item {\em Read control}. There are two dimensions to these actions: 
\begin{enumerate}
\item \emph{Wait}. This can let some dependent transaction $T'\in T_{dep}$ perform its conflicting write earlier than $T$'s read, resulting in $T'$\WR $T$. Otherwise, a dependency cycle may arise with $T$\RW $T'$, resulting in aborts.
\item \emph{Which version of data to read, including either committed or uncommitted version}. This amounts to choosing which location in the 
access list to insert the read, thereby affecting dependencies. Specifically, since a read returns the latest write $w$ before itself in the access list, there is a write-read dependency, $T'$\WR $T$, for every $T'$ whose write appears before this read in the list.
Additionally, a read also results in a set of read-write dependencies $T$\RW $T'$, for every $T'$ 
whose write appears after this read in the list.
\end{enumerate}

\item {\em Write control.} There are two dimensions to these actions:
\begin{enumerate}
    \item\emph{Wait}. The rationale for this action is similar to that for reads. 
    \item \emph{Whether or not to make this write visible to the future reads of other transactions.} The write is buffered if it is not exposed. Otherwise, this write as well as all of $T$'s previously buffered writes are made visible by appending them to the corresponding per-object access lists.
The cumulative way of exposing writes makes sense because otherwise, any transaction that has read this but not 
a previous write of $T$ would violate serializability and get aborted. Unlike a read, there's no flexibility to insert a write in any location but the end of the list; this is because we cannot allow a write to affect past reads. 
Exposing a write does not imply that uncommitted data will be read because transactions can choose to read committed versions only.  In terms of the resulting dependencies, exposing a write causes $T'$ \WW $T$ or $T'$ \RW $T$ for any $T'$ whose write or read appears before this write in the list. 
\end{enumerate}

\end{itemize}

\addtocounter{footnote}{-3}

\refstepcounter{footnote}
\label{fnt:1}

\vspace{0.1in}
\noindent \textbf{Available actions for validation.}  Actions in this category can control two aspects of validation:
\begin{itemize}
    \item {\em When to validate.} A transaction may validate its accesses at any time during execution, instead of only at commit time.  Early validation can abort a transaction quicker to reduce wasted work. 

    \item {\em How to validate.} The most precise form of validation is to explicitly check whether committing transaction $T$ would form dependency cycles with other committed transactions~\cite{mu2014extracting}.  However, such graph-based validation is expensive to implement for in-memory databases. A practical alternative is OCC-style validation~\cite{occ:kung81, silo:sosp13} which uses each transaction's physical commit-timestamp (cts) as its serialization order. Although such validation is conservative and has false aborts, it is fast. Prior work has also proposed validation based on logical commit-timestamps~\cite{yu2016tictoc}.
\end{itemize}

\subsection{Decomposing existing \cc algorithms}
We take a deep dive to study existing algorithms through the 
the lens of our framework. At a high level, existing algorithms differ from each other by the distinct combinations of action choices they have, even though their choices remain the same regardless of state.

As summarized in Table~\ref{table:classification}, traditional 2PL~\cite{2pl:gray75} and OCC~\cite{occ:kung81} algorithms both read the latest committed data.  OCC does not wait to perform any accesses nor does it expose the writes.  By contrast, 2PL exposes writes in order to block future conflicting accesses. We can approximate 2PL's blocking behavior by the action choice that makes transaction $T$ wait for all its dependent transactions $T_{dep}$ to commit before its data access. This approximation is slightly less aggressive than that of 2PL, which makes $T$ wait for $T'$ to commit if the current access {\em will} make $T$ dependent on $T'$.  We use the term 2PL$^*$ to refer to 2PL with this approximated blocking. Sundial~\cite{sundial:vldb18} handles write-write conflicts with 2PL and read-write conflicts with OCC; thus, it blocks write access until all its write dependencies $T_{wdep}$ commit and has no blocking for reads. As for validation, traditional algorithms do it only at commit time, except for 2PL whose deadlock detection or prevention mechanism can be viewed as a form of early validation done at every access.

Apart from traditional \cc, our framework also applies to a class of recently proposed algorithms including Callas RP~\cite{xie2015high}, IC3~\cite{wang2016scaling} and DRP~\cite{mu2019deferred}. These algorithms structure each transaction as a series of pieces~\cite{txchopping}, and try to pipeline the execution of these pieces to enhance performance under contention.  As shown in Table~\ref{table:classification}, unlike traditional \cc, they make a transaction's writes visible and allow reads of uncommitted data.  Furthermore, they make transaction $T$ wait before an access until $T$'s dependent transactions finish execution up to a certain point, determined by applying a static analysis of the transaction workload.

Although our design for learnable \cc (\S\ref{sec:design}) does not support MVCC, we can nevertheless examine MVCC algorithms using our framework. Table~\ref{table:classification} shows the actions made by MVTSO~\cite{database:bernstein}. Other MVCC algorithms~\cite{ports:postgres12, deuteronomy:cidr15, si:eurosys12} have similar actions but use different validation methods. Under MVTSO, a transaction reads the largest committed version smaller than its timestamp.  Writes are exposed so that future reads by transactions with larger timestamps will wait for this transaction to commit. MVTSO also performs a form of early-validation and aborts $T$ if there exists $T'\in T_{dep}$ such that $T'$\RW $T$ and $T'$ has been assigned a larger timestamp.

Not all \cc algorithms can be expressed by our framework.  In particular, our framework tracks dependencies and controls the interleaving of data access at runtime, and therefore cannot encode those \cc algorithms that pre-define dependencies according to some globally-agreed ordering prior to execution, e.g. Calvin~\cite{thomson2012calvin}, Granola~\cite{cowling2012granola}, Eris~\cite{eris:sosp17} and RoCoCo~\cite{mu2014extracting}. Moreover, our framework assumes that each access of the transaction is executed one after another by a single thread, and hence cannot encode algorithms like Bohm~\cite{Faleiro14vldb} that uses multiple threads to execute a single transaction.

\section{\sys Design}
\label{sec:design}

We design \sys according to the framework of \S\ref{sec:motivation}. The design consists of two parts: 1) a suitable policy space. 2) a training procedure to optimize the policy for a given workload.  This section describes the policy space. The next section (\S\ref{sec:training}) discusses training.

\vspace{-1mm}
\subsection{Overview}

\begin{figure}[t]
\centering
\includegraphics[scale = 0.4]{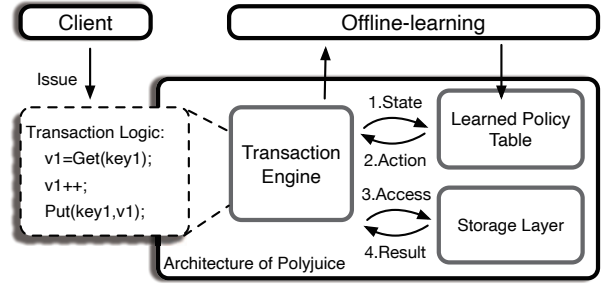}
\caption{System architecture. Before executing a specific data access in the transaction, \sys consults the learned actions in the policy table (step 1, 2). Then, \sys performs the access in the storage layer according to the actions.}
\label{fig:system-architecture}
\end{figure}

\noindent\textbf{System architecture.} \sys is a multi-core in-memory database. There is no multi-version support. For each data object, \sys stores the latest committed data as well as a per-object access list. The access list contains all uncommitted writes that have been made visible, as well as read accesses.  A transaction uses the access list to track the dependencies for each data access. \sys uses a pool of workers that run concurrently: each worker executes a transaction and commits it according to the learned \cc policy, which has been trained offline. Fig.~\ref{fig:system-architecture} shows \sys's system architecture.

\vspace{0.1in}
\noindent\textbf{Policy Representation.} As discussed in \S~\ref{sec:motivation}, we consider each learnable \cc algorithm as a policy function $p$ that maps from the {\em state space} ($S$) to the {\em action space} ($A$), $p: S\rightarrow A$.  Both the state and action space consists of a number of dimensions; the size of the state/action space is exponential w.r.t. the number of dimensions. 

We represent each policy function as a table: there are as many rows in the policy table as there are different states; there are as many columns as there are action dimensions. Such tabular representation is practical only if the state space is not too huge, which is the case in the workloads that we have studied. \S~\ref{sec:limit} discusses the limitation of large state space and potential solutions.

\begin{figure}[t]
\centering
\includegraphics[scale = 0.32]{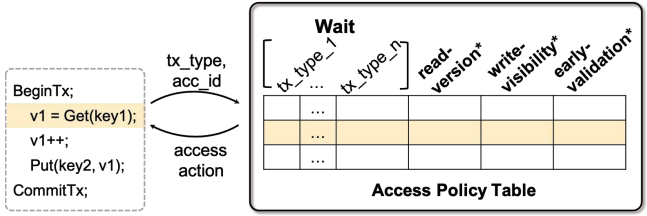}
\caption{Policy table (* indicates a binary field).}
\label{fig:policytable}
\end{figure}

For a given \cc policy table, a cell $c_{i,j}$ at row $i$ and column $j$ indicates that for the access with execution context (state) $i$, the system should take the action given by cell $c_{i,j}$ for action type (a.k.a. dimension) $j$.  In \sys, each cell contains either a binary number for a binary action (e.g. whether to make writes visible or not), or an integer for a multi-valued action (e.g. how to wait for dependent transactions). Fig.~\ref{fig:policytable} shows the \cc policy table; details on its rows and columns are explained in \S\ref{subsec:state} and~\ref{subsec:action}. \sys learns the backoff time for retrying aborted transactions separately (\S\ref{subsec:backoff}).

\vspace{0.1in}
\noindent\textbf{Policy-based Execution.}  In \sys, the database is given the learned policy table with which to perform concurrency control.  To execute a transaction according to the policy, \sys 
looks up in the policy table at each data access to determine the corresponding set of actions.  Some of these actions 
are to be performed prior to the data access, e.g. whether and how long to wait, while others are to be done after the access, e.g. making a write potentially visible by appending it to the access list. 
After finishing execution, \sys commits a transaction after  performing the final validation to ensure serializability (\S~\ref{subsec:validate}).

\subsection{\cc policy: state space}
\label{subsec:state}

The term {\em state} is from the RL literature. In our case, state can be viewed as the execution context of the current data access.  Ideally, the state space should be able to distinguish execution contexts that are best served by different actions. It should also be limited in size so that the resulting policy table is not too huge and can be
searched efficiently during training. 

\sys's state space contains the following information:
\begin{enumerate}[leftmargin=*,noitemsep,nolistsep]
    \item The type of the transaction being executed. For a given workload whose transactions are specified in stored procedures, the type can be identified 
    by the stored procedure name.
    \item Which access of the transaction is being executed.  We use an integer access-id to identify each access. Access-id is determined by the static code location that invokes the access. Using static information for access-id provides a good trade-off: it can discriminate most accesses while avoiding blowing up the state space. 
\end{enumerate}

It is tempting to include other useful information, such as which type of access (read/write/commit) 
and which data table is being executed. Interestingly, for most workloads, both of these can be uniquely determined 
by the access-id and thus we omit them from the state space. We have also experimented with adding the contention level of the accessed data to the state space.
However, we found that doing so only benefited a few contrived micro-benchmarks.  In practical workloads including TPC-C and TPC-E, distinguishing transaction type and access-id is sufficient to capture the main contention characteristics. Even for artificial workloads, it is difficult to find a scenario where including contention level results in noticeable performance improvements.  Including contention level makes it possible to differentiate accesses with the same access id. However, \sys's wait action (\S~\ref{subsec:action}) cannot take advantage of such differentiation.

\vspace{0.1in}   
\noindent\textbf{Size of state space (a.k.a. number of different states).} The state space size determines the number of rows in the \cc policy table. 
Let $n$ be the number of different transaction types in a workload, and $d_1$, $d_2$, ..., $d_n$ be the number of static data accesses for transaction of type $1, 2, ..., n$. Then the state space size (i.e. number of different states) is: $d_1+d_2+...+d_n$.  

\subsection{\cc policy: action space}
\label{subsec:action}

\sys's action space contains knobs in two categories: interleaving control and validation. 

\vspace{0.1in}
\noindent\textbf{Supported actions for interleaving control.} There are three classes:
\begin{itemize}
    \item {\em Wait}. This action is invoked {\em before} a read or write.  How to specify how long the wait should be?  A naive design is to use absolute time intervals, but this makes the wait action sensitive to execution time variations, resulting in fragile policies.
    
    \hspace{\parindent} Since the goal of waiting is to let another potentially conflicting transaction to go ahead with its data access, we quantify how long transaction $T$ should wait by how much progress the transactions that $T$ depends on have made so far. This design is inspired by existing protocols like Callas RP~\cite{xie2015high} and others~\cite{wang2016scaling, mu2019deferred}.  More concretely, we group transactions by type, and measure the execution progress of a transaction type by access-id. 
    The special value NO\_WAIT indicates no waiting. Suppose the wait action for transaction type $X$ has access-id $a$, then transaction $T$ must wait for all $T$'s dependent transactions of type $X$ to finish execution up to and including $a$. 
    For a workload with $n$ different types of transactions, the wait action consists of $n$ access-ids, one for each transaction type.  
    
    \item {\em Read-version}. This action has a binary choice: CLEAN\_READ for reading the latest committed version, DIRTY\_READ for reading the latest uncommitted (but visible) version.  Although there may be more than one uncommitted copy of data, there is no point in reading an earlier version because doing so would result in more dependencies and higher abort likelihood. 
    
    \item {\em Write-visibility}. This action is invoked {\em after} a write access and is also binary: PRIVATE keeps the write in the private buffer, PUBLIC makes all private writes buffered so far visible by appending them to the access list.

\end{itemize}

\vspace{0.1in}
\noindent\textbf{Supported actions for validation.}
Validation always happens before commit (\S~\ref{subsec:validate}). \sys also supports the action of early-validation, which can occur {\em after} any read/write.   If it's set, this binary-valued action checks if the reads and writes done since the last validation have violated serializability. Earlier accesses, which have passed previous early-validation, are likely to have already been serialized and thus not checked.  Early-validation does not guarantee correctness but avoids wasting work by detecting non-serializable access early.

\sys supports the wait action before early-validation. The encoding of the wait action is the same as that for reads/writes. To reduce the action space, we consolidate the two kinds of wait actions into one. In particular, \sys uses the wait action corresponding to the next access-id if early-validation is enabled for the current access-id. 

Upon failing early-validation, \sys retries the transaction from the point of its last successful validation. In order to reduce the cost of the failed validation, we defer appending reads and visible-writes to their corresponding access lists until a successful early-validation.  Otherwise, 
failing early-validation means having to remove previously appended reads/writes from access lists, and to abort transactions that have read those discarded writes.  Conceptually, we can separate the decision of early validation from that of appending reads/writes to access lists.  However, in our experience, doing so complicates the implementation without improving the final learned \cc performance.

\vspace{0.1in}
\noindent\textbf{Size of action space (a.k.a. number of different action choice combinations per state).} Let $n$ be the number of different transaction types in a workload, and $d_1$, $d_2$, ..., $d_n$ be the number of static data accesses for transaction of type $1, 2, ..., n$. Then the number of different action choice combinations can be calculated as: $d_1*d_2*...*d_n ($wait\ choices$) * 2 ($read-version$) * 2 ($write-visibility$) * 2 ($early-validation$)$.

\subsection{Validation for correctness}
\label{subsec:validate}

\sys uses an OCC-style physical timestamp-based validation in the final commit phase to ensure correctness. To commit a transaction $T$ with validation, a worker takes 4 steps: 1) it waits for all $T$'s dependent transactions to commit (or abort). 2) it locks each record in $T$'s writeset 3) it validates each record in the readset by checking two conditions; whether the version-id of the current committed version in the database is different from that kept in the readset, and whether the record is being locked by another transaction. If either condition is true, $T$ is aborted. 4) if validation succeeds, it applies $T$'s writes to the database along with their version-ids, and releases the locks.

Our validation algorithm is identical to that of Silo~\cite{silo:sosp13} except for two additional mechanisms which are crucial for correctness. First, we use a unique 
version-id for committed as well as uncommitted versions, because the latter may be read from the access list. Second, we add the additional first step of waiting for $T$'s dependent transactions to finish committing.  
We provide a brief correctness argument here. 
\ifx \arxiv\mode
A more detailed proof is in the Appendix. 
\fi
\ifx \camera\mode
A more detailed proof is in the Appendix of the extended version~\cite{extended}.
\fi
We argue the correctness of \sys by reduction to Silo: if \sys commits a transaction, then Silo would also commit it. According to step-1, \sys ensures that if a transaction $T$ is committed successfully, then before $T$ starts the validation, all of its dependent transactions (e.g. $T_{dep}$) have been committed. This allows us to prove that executing $T$ is equivalent to executing another hypothetical transaction $T'$ which starts execution after all transactions in $T_{dep}$ commit, reads/writes the same data as $T$, and starts validation at the same time as when $T$ starts its validation. 
Therefore, if $T$ passes the validation in \sys, $T'$ can pass the validation of Silo and successfully commit itself.

\subsection{Learning backoff time}
\label{subsec:backoff}
Separate from the \cc algorithm, it is also important for performance to use an appropriate backoff time for retrying an aborted transaction. Existing systems, e.g. Silo, use simple binary exponential backoff which doubles the backoff time with each failed attempt.  This simple strategy is inadequate as it often results in backoff times that are too short in the first couple of retries but too large after several successive retries.  Furthermore, this strategy does not distinguish between different transaction types when adjusting backoff times. This is suboptimal: intuitively, one can increase the backoff time more aggressively for a transaction type more prone to contention.

For learning the backoff time, \sys uses a separate backoff policy table. The rows (a.k.a. state space) of this table enumerate 3 dimensions: 1) the transaction type 2) the status of the current execution (commit or abort). 3) the number of aborted attempts prior to the current execution with a fixed cutoff: our current implementation uses 0, 1 or 2 to indicate whether there has been 0, 1 or 2+ aborts so far. The action space of the backoff policy table is inspired by recent work on learnable congestion control in  networking~\cite{jay2019deep}. Specifically, a worker adjusts the backoff time for each transaction type multiplicatively whenever it commits/aborts a transaction:
$$
\mathit{backoff}=
\begin{cases}
\mathit{backoff} \times (1 + \alpha_{t,i, aborted}), & abort\\
\mathit{backoff} / (1 + \alpha_{t,i,committed}), & commit
\end{cases}
$$
In the above equations, $\alpha_{t, i, committed}$ or $\alpha_{t, i, aborted}$ is the learned parameter (a.k.a. action) in the policy table for transaction type $t$, number of prior aborted attempts $i$ and execution status $committed$ or $aborted$. To enable easier training, we use bounded discrete values for $\alpha$.  In particular, $\alpha$ can be zero, resulting in unchanged backoff time.

\section{Training Policies}
\label{sec:training}
\begin{paragraph}
{\bf Overview} The policy space discussed in \S\ref{sec:design} is exponentially large: there are $a^s$ different policies, where $s$ is the number of different states and $a$ is the number of different actions per state. The goal of training is to efficiently search for a good policy for a given workload.

\sys performs training offline. During regular execution, \sys logs executed
transactions together with their inputs.  Using a separate training machine, \sys emulates the target workload by reissuing transactions with their logged inputs. 
We measure a policy's commit throughput under the emulated workload.

{\sys} uses Evolutionary Algorithm (EA) for training.  We have also explored the policy-gradient method from the RL literature (\S\ref{subsec:rl}). Despite EA's simplicity, we have found it to be more efficient than the alternative (\S\ref{subsec:training}). 
\end{paragraph}

\subsection{Training using Evolutionary Algorithm}
\label{subsec:basic_ea}

EA is an optimization approach to search for a solution with good fitness by evolving a population of
individuals via 
nature-inspired mechanisms such as crossover, mutation, and selection~\cite{davis1991handbook, goldberg1988genetic, holland1992adaptation}.
In \sys, the fitness of an individual (aka a candidate policy) corresponds to the policy's commit throughput under the given workload.

EA starts by initializing the first generation of the population. The size of 
the population for each iteration is a configurable hyperparameter. To create a new children generation, EA performs mutation on the policies (including \cc and backoff policies) of the current generation (parents). 
It then evaluates the ``fitness'' of each mutated child by measuring its throughput. Finally, EA selects $N$ individuals according to their fitness to survive to the next generation.

\vspace{0.1in}
\noindent\textbf{Mutation.} EA mutates each cell of a parent's \cc and backoff policy table independently with probability $p$. If the cell corresponds to a binary choice such as read-version or write-visibility, the mutation flips the choice.
If the cell corresponds to an integer choice (e.g. any of the wait actions), the mutation varies the integer value by some distance uniformly sampled from the interval $[-\lambda, \lambda]$.
The mutated integer is clipped to always lie within the valid range.  The initial values of mutation probability ($p$) and mutation interval ($\lambda$) are configurable hyperparameters.  We decrease $p$ and  $\lambda$ gradually as the training progresses to facilitate convergence. This is akin to the decrease in learning rate in gradient descent methods or the gradual reduction of temperature in simulated annealing. 
 
Crossover, another popular EA mechanism, is not effective in our context.  Crossover endows a child's policy with some rows from one parent and some rows from the other parent.  Unfortunately, such a child is likely to perform worse than either of its parents.  This is because, in most good policies, the wait actions of different rows are not independent but highly correlated. Thus, mixing the rows of different policies often results in worse performance.

At the end of each iteration, EA chooses $N$ individuals with the best performance from the current population to survive to the next iteration.  In our experiments, this simple selection mechanism trains faster than tournament selection~\cite{davis1991handbook, goldberg1988genetic, holland1992adaptation}.

\vspace{0.1in}
\noindent\textbf{Warm start.} Instead of using all random policies, we seed the initial population with several known good policies, 
including OCC, 2PL$^*$, and Callas RP/IC3.  These policies are likely not optimal for the given workload, but they provide some good initial policies to give EA a ``warm start'' in training.

\subsection{Alternative training method}
\label{subsec:rl}

Some recent works have used RL training methods to solve systems problems such as task scheduling~\cite{mao:sigcomm19}, adaptive video streaming~\cite{pensieve:sigcomm17}, multi-GPU dataflow systems~\cite{deviceplacement,mirhoseini2018hierarchical}, congestion control~\cite{jay2019deep}, etc.
We have experimented with the policy-gradient method for training a parameterized stochastic policy~\cite{reinforce}. More concretely, we parameterize the policy table by representing each table cell using one or a set of parameters to denote the probability distribution of the action values. Suppose the cell at coordinate $i,j$ corresponds to some action with $M$ possible choices, we use $M$ parameters, $p_{i,j}^0, p_{i,j}^1, ... p_{i,j}^{M-1}$, which are fed into a softmax function to denote the probability distribution of $M$ choices.  

For training, each iteration samples a batch of policies according to the probability distribution specified by the current table parameters.  We measure the throughput of each sampled policy and use it as the ``reward'' in RL. Policy gradient maximizes the expected reward by performing gradient descent~\cite{reinforce}. Our way of applying policy gradient is inspired by~\cite{proxylessnas}. We compare RL- and EA-based training in \S\ref{subsec:training}.

\subsection{Training for real-world deployments}
\label{subsec:deployment}

Since \sys relies on offline training to optimize its policy for a 
specific workload, this raises the question of how to use it in the real-world with changing workloads.  We acknowledge that \sys is not suitable for very dynamic and unpredictable workloads.  However, we observe that many real-world workloads are fairly {\em predictable} on a day-to-day basis after analyzing the trace of an  e-commerce website. This has motivated us to suggest the following deployment strategy for \sys.


\vspace{0.1in}
\textbf{Optimize for the peak workload.} Real-world systems are provisioned for the anticipated peak workload. Hence, our goal is to use \sys to improve  commit throughput during the peak time, in which the server receives the most requests in a day. 
There is no need to optimize for non-peak workloads because an under-utilized database is not a bottleneck for application performance.  Therefore, we only need to train the policy tailored to the peak workload, and run the same policy during non-peak times as well.

\vspace{0.1in}
\textbf{Predict and retrain.} Our analysis of the real-world trace shows that one can predict tomorrow's peak workload characteristics using the statistics gathered from today's peak workload (\S\ref{subsec:dynamic-workload}). Given this observation, one can collect the trace of the peak hour today, retrain the policy based on the trace, and run this policy for tomorrow. Doing so naively requires \sys to retrain the policy every day.  We can defer retraining if the predicted peak workload does not differ significantly from the one targeted by the current policy.  Our analysis of the real-world trace shows that the peak workload can remain stable for many days after a significant change. Hence, deferral can greatly decrease the number of retraining times. One is right to be concerned that deferred training and prediction errors can result in running a policy optimized for a different workload than the actual one happening. We also evaluate the effect of this discrepancy in \S\ref{subsec:dynamic-workload}.

\section{Implementation}
\label{sec:impl}

We implemented \sys in C++ using the codebase of Silo~\cite{silo:sosp13} by replacing Silo's concurrency control mechanism with \sys's policy-based algorithm.
We implemented \sys's offline training separately in Python (and RL-based training in TensorFlow).  The result of training is the policy table, which is written to disk as a file and later loaded into memory by the C++ database. Each worker thread in the C++ database maintains a pointer to the in-memory policy table. When switching the policy, we reset the policy pointer in each worker thread.  \sys doesn't need to atomically switch the policy pointers of all threads. This is because \sys's validation procedure can ensure correctness regardless of the policies used during execution.

Like Silo, transaction logic is written in C++ using a few API calls (e.g. Get/Put/CommitTx). 
Each Get/Put/CommitTx API call's access-id is its corresponding sort order based on the API invocation's line number. For range queries, our current prototype reuses Silo's existing mechanism which always reads the committed value.

\ifx \arxiv\mode
The pseudocode of how \sys executes a transaction according to the policy is included in the Appendix.
\fi
\ifx \camera\mode
The pseudocode of how \sys executes a transaction according to the policy is included in the Appendix of the extended version~\cite{extended}.
\fi

\section{Evaluation}
\label{sec:evaluation}

\begin{figure*}
\begin{minipage}{.74\textwidth}
\centering
\subcaptionbox{High Contention \label{fig:tpcc-performance-high-contention}}{\includegraphics[width=.32\linewidth]{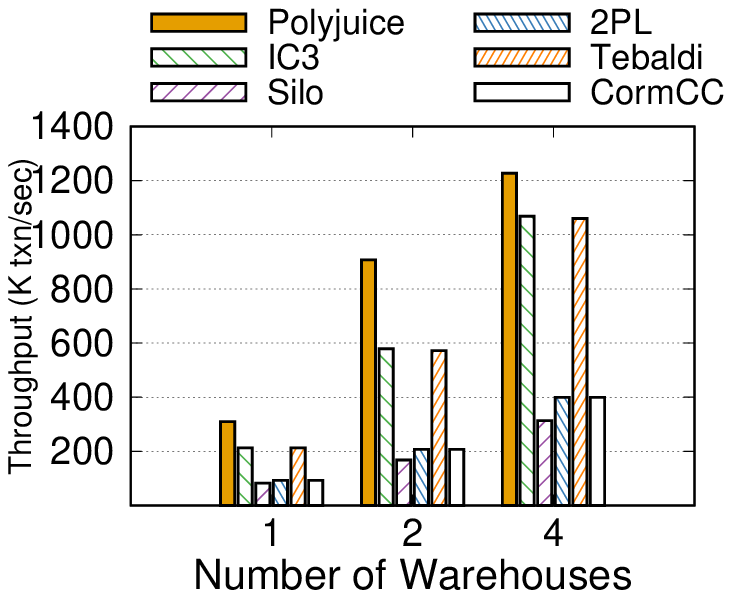}}
\subcaptionbox{Moderate to Low Contention \label{fig:tpcc-performance-low-contention}}{\includegraphics[width=.32\linewidth]{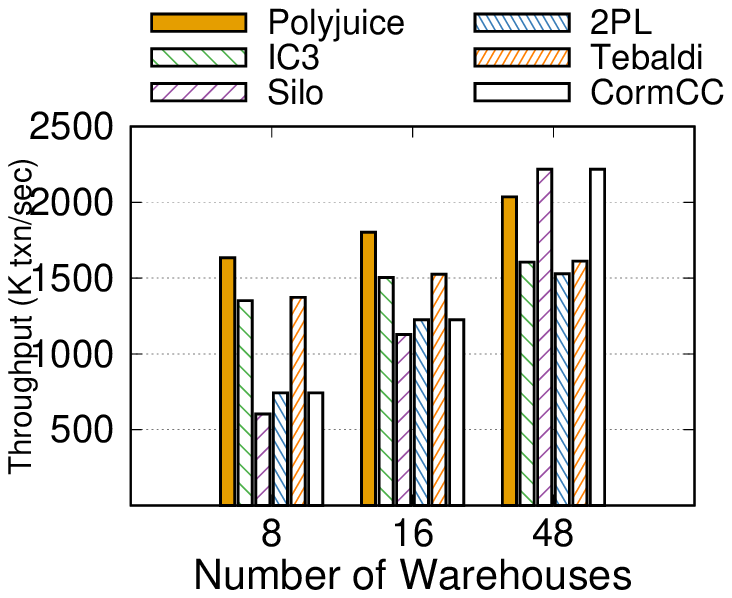}}
\subcaptionbox{Scalability (1 WH) \label{fig:tpcc-scalability-high-contention}}{\includegraphics[width=.32\linewidth]{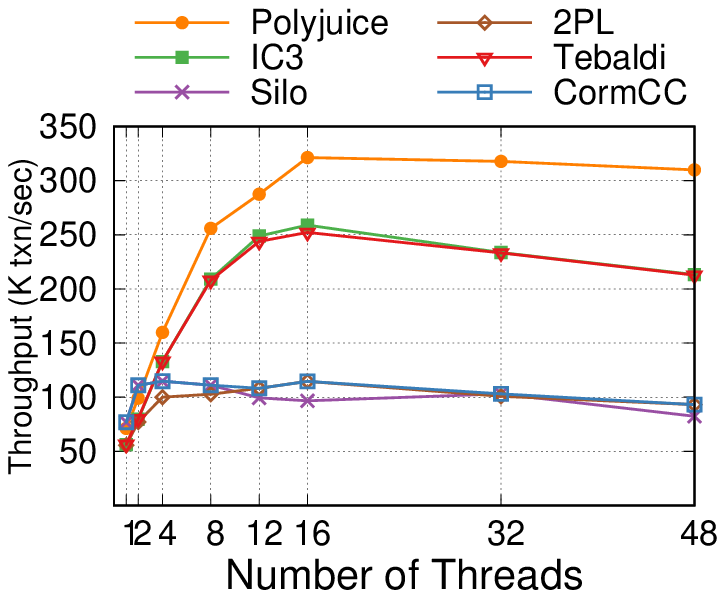}}
\caption{TPC-C Performance and Scalability}
\label{fig:TPCC}
\end{minipage}
\begin{minipage}{0.24\textwidth}
\includegraphics[width=\linewidth]{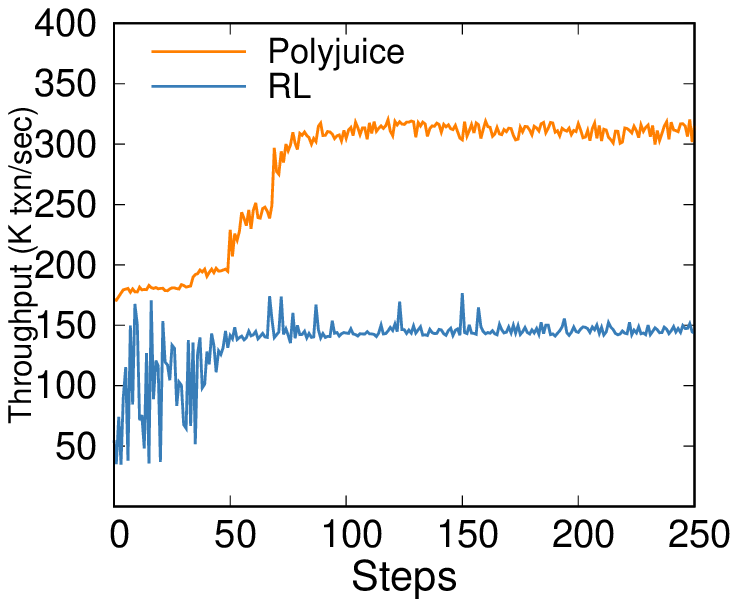}
\caption{EA v.s. RL}
\label{fig:training-RL}
\end{minipage}
\end{figure*}

\begin{table*}
\centering
\begin{adjustbox}{max width=\textwidth}
\begin{tabular}{cccccccc}
\hline
& & \sys& IC3& Tebaldi& Silo& 2PL& CormCC\\
\hline
\multirow{3}*{Latency($\mu$s)}

&Neworder& 163/151/179/245&	251/246/296/345&	246/240/291/354&	1084/20/62/263&	450/31/49/178&	450/31/49/178\\

&Payment& 163/151/181/252&	247/242/291/340&	242/236/285/348&	6/4/9/24&	658/19/97/1554&	658/19/97/1554\\

AVG / P50 / P90 / P99&Delivery& 172/167/194/269&	156/152/177/223&	155/151/175/208&	108/101/120/248&	183/145/279/621&	183/145/279/621\\

\hline
\end{tabular}
\end{adjustbox}
\caption{Latency for each transaction type in TPC-C with 1 warehouse and 48 threads}
\label{table: tpcc-breakdown}
\end{table*}










\subsection{Experimental setup}
\label{subsec:setup}
{\bf Hardware}. Our experiments are conducted on a 56-core Intel machine with 2 NUMA nodes. Each NUMA node has 28 cores (Xeon Gold 6238R 2.20GHz) and 188GB memory.

\noindent {\bf Workloads}. We use three benchmarks, TPC-C\cite{TPC-C}, TPC-E\cite{TPC-E}, and a micro-benchmark with ten types of transactions. In our experiments, each worker retries an aborted transaction indefinitely until success, to ensure that committed transactions adhere to the workload's specified mix ratio of different transaction types. If we had not done this and let a worker give up an aborted transaction and start a new one with a different type, we would incorrectly learn a policy that intentionally aborts some transaction types to maximize aggregate throughput.

\noindent {\bf Baselines for comparison }. We compare \sys with five existing algorithms: OCC (Silo)\cite{silo:sosp13}, 2PL\cite{2pl:gray75}, IC3\cite{wang2016scaling}, Tebaldi\cite{su2017bringing} and CormCC\cite{tang2018toward}. For Silo and IC3, we use the authors' source code.  For Tebaldi and CormCC, we simulate them in our codebase to provide an apples to apples comparison. For 2PL, we implement it in Silo's codebase with an optimized WAIT-DIE mechanism. The optimization avoids aborts if locks are acquired following a global order, as is the case with our TPC-C and microbenchmark.

\noindent {\bf Methodology}. For the training, we use 300 iterations by default.  After each iteration, we pick 8 policies from the current population. For each of them, we generate 
another 4 children policies and add them to the selection pool. Therefore, there are a total of $8*5=40$
policies at each iteration. To evaluate the performance of the learned policy as well as other baseline algorithms, we run the workload five times, with each run taking 30 seconds. By default, the graphs show the median.

\subsection{TPC-C}
\label{subsec:tpcc}

For the TPC-C benchmark, we evaluate the three read-write 
transactions only, as the remaining two read-only transactions can be processed with the snapshot mechanism derived from Silo.  We vary the number of warehouses in the benchmark to change the level of contention.

By default, we use the 3-layer configuration for Tebaldi, which divides the read-write transactions into two groups (NewOrder, Payment vs. Delivery) isolated by 2PL~\cite{su2017bringing}. Tebaldi's 2-layer configuration puts all read-write transactions into the same group, which is the same as IC3. We simulate CormCC according to its paper\cite{tang2018toward}.  In particular, we partition the workload by warehouse so that all accesses to the same warehouse are protected by the same \cc. 
Moreover, as all warehouses are inter-changeable in our benchmark, all partitions should also use the same \cc protocol. 
Based on this observation, 
we measure the performance of 2PL and OCC, and pick the one with the better performance as the \cc protocol for each partition. 

\noindent \textbf{Throughput.} 
\Cref{fig:tpcc-performance-high-contention} and \ref{fig:tpcc-performance-low-contention} show the throughput of various algorithms  
with 48 threads under different contention levels. \Cref{fig:tpcc-performance-high-contention} gives the throughput under 
high contention. \sys achieves significant
performance improvements. Specifically, with two warehouses, its throughput
reaches 907K TPS, which is more than $1.5\times$ of other algorithms.
IC3 and Tebaldi have higher throughput than other existing algorithms 
because they can exploit a form of ``pipelined'' execution. Both have the same throughput, which differs from the original paper, as we 
disable their manual optimization for commutativity and uniqueness. 
Compared with IC3 and Tebaldi, 
\sys achieves 56\% improvement because of two factors:  First, it can avoid unnecessary waiting because it uses the runtime information 
to infer the \cc action, while IC3 only leverages the 
static information. Second, \sys can 
either read dirty or clean versions of data. This flexibility 
enables it to achieve more efficient interleavings.
We provide a detailed analysis with an example in \S~\ref{subsec:case-study}.

\Cref{fig:tpcc-performance-low-contention} shows the throughput under moderate and low contention. \sys outperforms the others for 8 and 16 warehouses. 
For 48 warehouses, in which each worker corresponds to its local warehouse, \sys
is slightly slower (8\%) than Silo, even though \sys learns the same policy 
as Silo. This is because \sys needs to maintain 
additional metadata in each tuple, which affects the cache locality. 

\noindent \textbf{Scalability.} \Cref{fig:tpcc-scalability-high-contention} shows the scalability of \sys under high contention (1 warehouse). 
\sys has the same scalability as IC3 and Tebaldi, which can scale to 16 threads. Compared with them, Silo and 2PL do not scale beyond 
four threads because they cannot exploit parallelism under  high contention.
CormCC also has the scalability 
issue because it is limited by the protocols (2PL and OCC) it uses. 

\noindent \textbf{Performance of each transaction type}.
We also study the throughput and latency for each type of read-write transaction 
with 1-warehouse and 48 threads (\Cref{table: tpcc-breakdown}). 
For \sys, the throughput of each type is 132K (NewOrder), 126K (Payment) and 11K (Delivery) TPS, which follows TPC-C specified ratio (45:43:4) very closely. This is because each worker retries an aborted transaction infinitely until it succeeds before starting a new transaction. Therefore, the ratio of the per-type commit throughput is exactly the same as how each worker generates these types. For the latency of NewOrder, 
\sys has higher P99 latency than 2PL, but lower latency than 
Silo, IC3 and Tebaldi. 
For Delivery, the outcome is flipped: \sys has lower P99 latency than 2PL, but higher latency than Silo, IC3 and Tebaldi.
For Payment, \sys has lower P99 latency than IC3, Tebaldi and 2PL.

\begin{figure}
\begin{subfigure}[t]{0.22\textwidth}
\centering
\includegraphics[width=\textwidth]{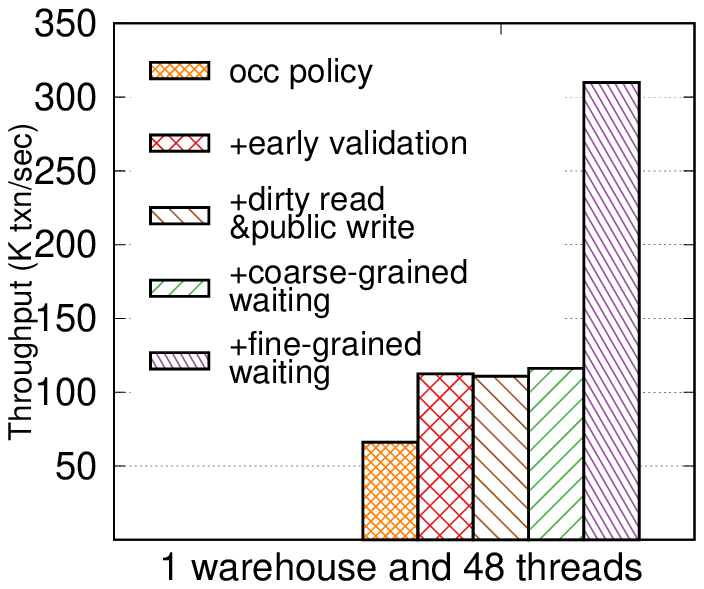}
\caption{High Contention}
\label{fig:factor-analysis-1wh}
\end{subfigure}
\hspace{0.05cm}
\begin{subfigure}[t]{0.22\textwidth}
\centering
\includegraphics[width=\textwidth]{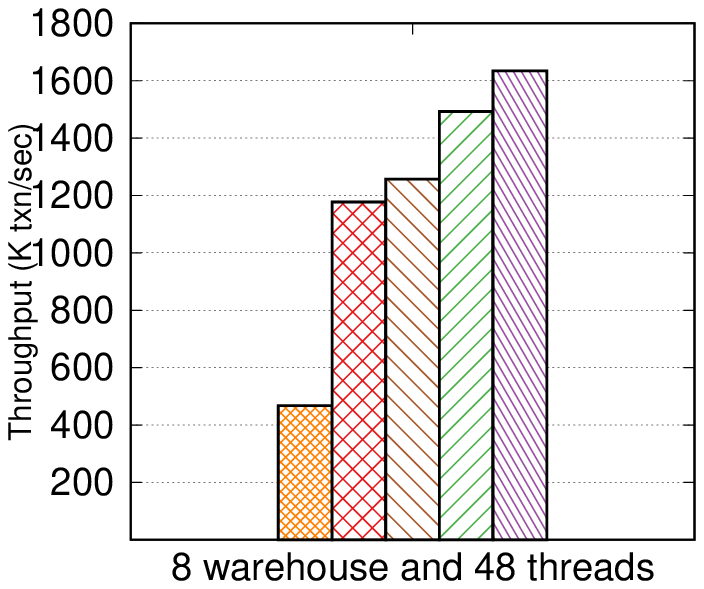}
\caption{Moderate Contention}
\label{fig:factor-analysis-8wh}
\end{subfigure}
\caption{Factor Analysis On TPC-C Benchmark}
\label{fig:factor}
\end{figure}

\noindent \textbf{Factor analysis}. To better understand the advantages of \sys, 
we perform a factor analysis to examine the benefits of different actions. 
We start with a policy including only the actions of OCC (Table \ref{table:classification}). Then, we gradually add other actions into the action space and measure the 
performance improvements. We classify the waiting actions into 
coarse-grained waiting and fine-grained waiting. The former means the actions of waiting 
for the dependent transaction to commit and learning the backoff. The latter 
refers to waiting for a certain access of the dependent transaction to finish. 

\Cref{fig:factor-analysis-1wh} and \ref{fig:factor-analysis-8wh} show the factor analysis result with 1 and 8 warehouses. For the 1-warehouse workload, adding ``early validation'' into the 
action space can improve the performance by 70\%, because it can detect the 
conflicts earlier and reduce the retry cost. 
\sys gets a performance boost after applying fine-grained waiting actions (116K to 309K TPS) due to full exploitation of the potential parallelism. However,  each action has a different effect factor with different workloads. For the 8-warehouse workload, adding ``early validation'' achieves larger 
improvement (467K to 1177K TPS) than others.

\begin{figure}
\begin{subfigure}[t]{0.22\textwidth}
\centering
\includegraphics[width=\textwidth]{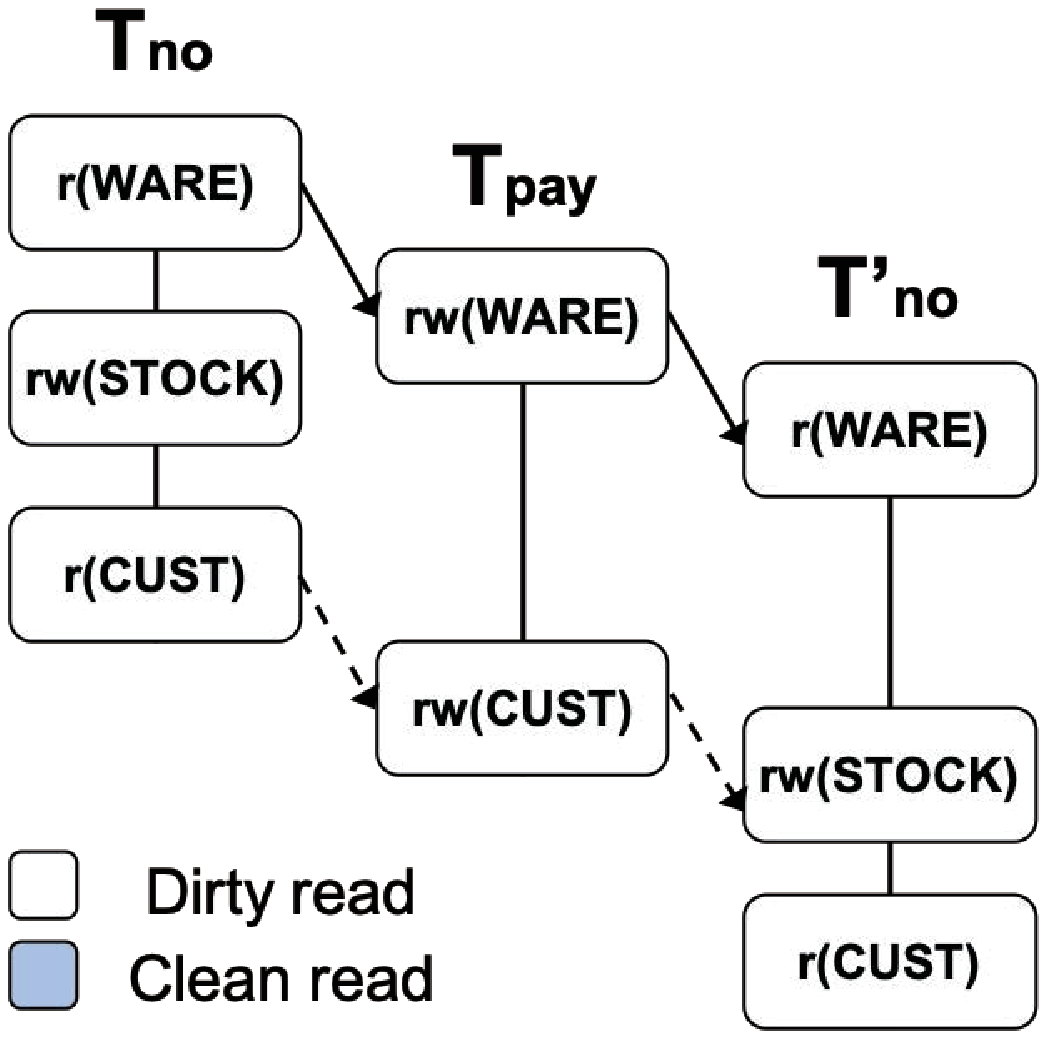}
\caption{IC3 interleaving.}
\label{fig:ic3-tpcc-interleaving}
\end{subfigure}
\hspace{0.5cm}
\begin{subfigure}[t]{0.22\textwidth}
\centering
\includegraphics[width=\textwidth]{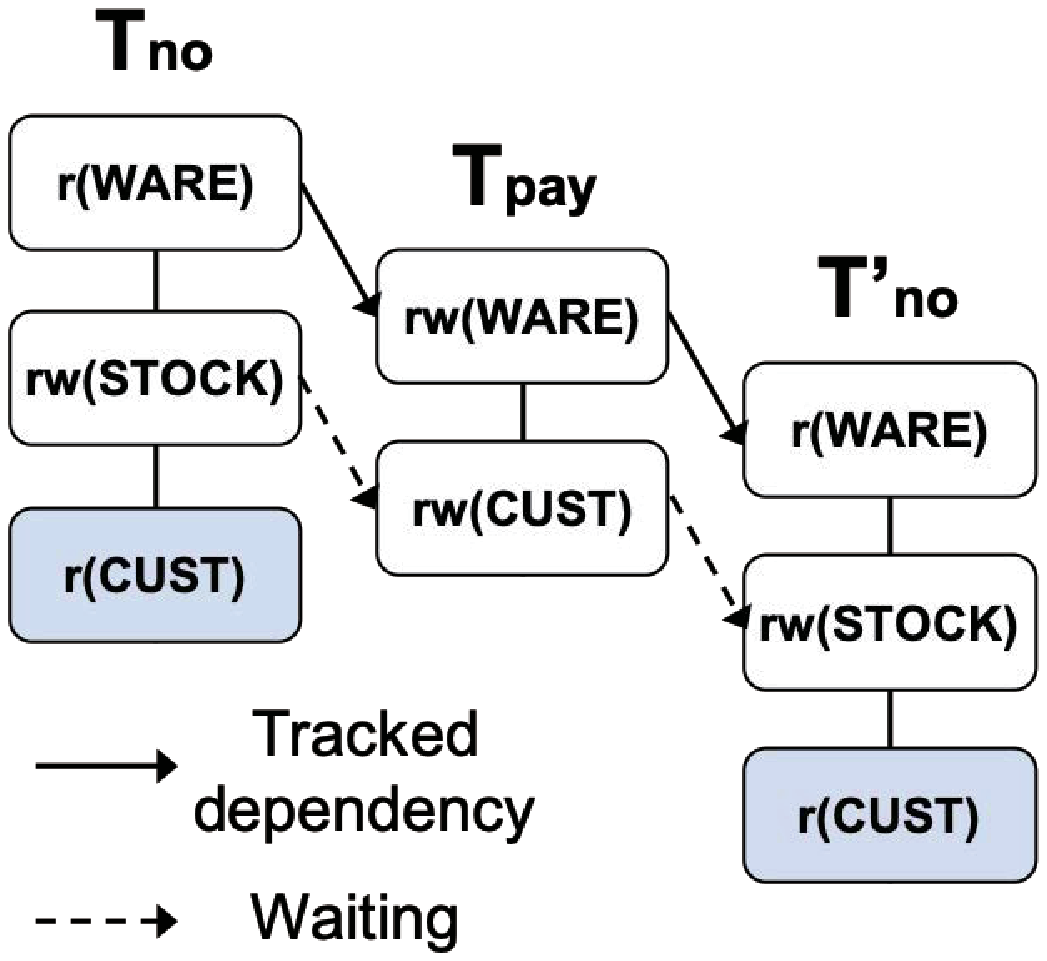}
\caption{\sys interleaving.}
\label{fig:chamcc-tpcc-interleaving}
\end{subfigure}
\caption{\sys's learned policy results in a more efficient interleaving for TPC-C than IC3.}
\label{fig:case-study}
\end{figure}

\begin{figure}
\begin{subfigure}[t]{0.22\textwidth}
\centering
\includegraphics[width=\textwidth]{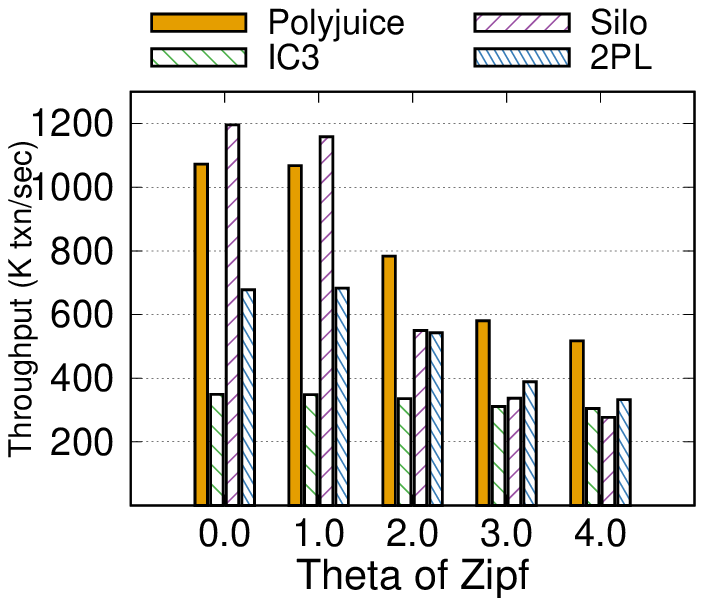}
\caption{Throughput}
\label{fig:tpce-performance}
\end{subfigure}
\hspace{0.05cm}
\begin{subfigure}[t]{0.22\textwidth}
\centering
\includegraphics[width=\textwidth]{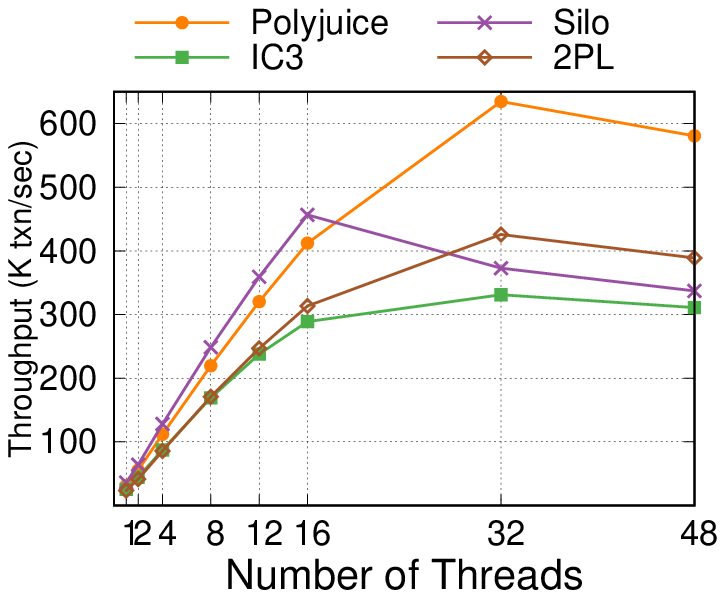}
\caption{Scalability ($\theta$ is 3.0)}
\label{fig:tpce-scalability}
\end{subfigure}
\caption{TPC-E Performance and Scalability}
\end{figure}

\subsection{A case study of learned policy}
\label{subsec:case-study}

We analyze an example learned policy to understand how it outperforms existing \cc algorithms.

\Cref{fig:case-study} shows an example of how IC3 and our learned policy mediate the data access of 3 concurrently executing transactions: $T_{no}$ (NewOrder), $T_{pay}$ (Payment) and $T_{no}'$ (NewOrder). All three access the same warehouse. \Cref{fig:case-study} shows a few crucial data accesses for each transaction: For NewOrder transactions ($T_{no}$, $T'_{no}$), these accesses are: read from WAREHOUSE table (r(WARE)), followed by an update to STOCK table (rw(STOCK)), and finally read from CUSTOMER table (r(CUST)). The crucial accesses of Payment ($T_{pay}$) are: update to WAREHOUSE (rw(WARE)) and update to CUSTOMER (rw(CUST)). 

The three transactions conflict because they access the same record in WAREHOUSE.  \Cref{fig:case-study} shows a specific dependency pattern that can arise from their WAREHOUSE access, $T_{no, r(WARE)}\rightarrow T_{pay, rw(WARE)} \rightarrow T'_{no, r(WARE)}$ as all WAREHOUSE accesses use dirty reads. As shown in \Cref{fig:ic3-tpcc-interleaving}, to avoid the dependency cycle, IC3 makes $T_{pay}$'s read of CUSTOMER wait for $T_{no}$'s CUSTOMER update to finish. This is because IC3 always uses dirty reads, so $T_{pay, rw(CUST)}$ must be ordered after $T_{no, r(CUST)}$ in accordance with their WAREHOUSE access' ordering. IC3 also makes $T_{no}'$ STOCK update wait for $T_{pay}$'s CUSTOMER update, even though these two access different tables. This is because IC3 only tracks the immediate dependency: by waiting for $T_{pay}$'s CUSTOMER update, it ensures that $T_{no}$ and $T_{no}'$ will not form a dependency cycle even though $T'_{no}$ is not aware of the transitively dependent $T_{no}$.

\Cref{fig:chamcc-tpcc-interleaving} shows the interleaving obtained by \sys, which is more efficient.
Unlike IC3, the learned policy makes $T_{pay}$'s CUSTOMER update wait for $T_{no}$'s STOCK access which is earlier than $T_{no, r(CUST)}$.  This shorter wait works because the learned policy also makes $T_{no}$'s CUSTOMER read a committed version, which helps avoid the conflict between $T_{no, r(CUST)}$ and $T_{pay, rw(CUST)}$. This is in contrast to IC3, which makes $T_{no, r(CUST)}$ perform a dirty read.
The learned policy still makes $T_{no}$'s STOCK update wait for $T_{pay}$'s CUSTOMER update like IC3 does, but the overall interleaving is more efficient.

Apart from IC3, neither CormCC nor Tebaldi can exploit this interleaving. CormCC does not allow dirty reads. Tebaldi uses the same action (either dirty or clean read) for all accesses within a transaction. \Cref{fig:chamcc-tpcc-interleaving}'s interleaving requires using dirty reads for NewOrder's WAREHOUSE access and clean reads for CUSTOMER access. 

\begin{figure}[t]
\begin{minipage}[t]{0.22\textwidth}
\centering
\includegraphics[width=\textwidth]{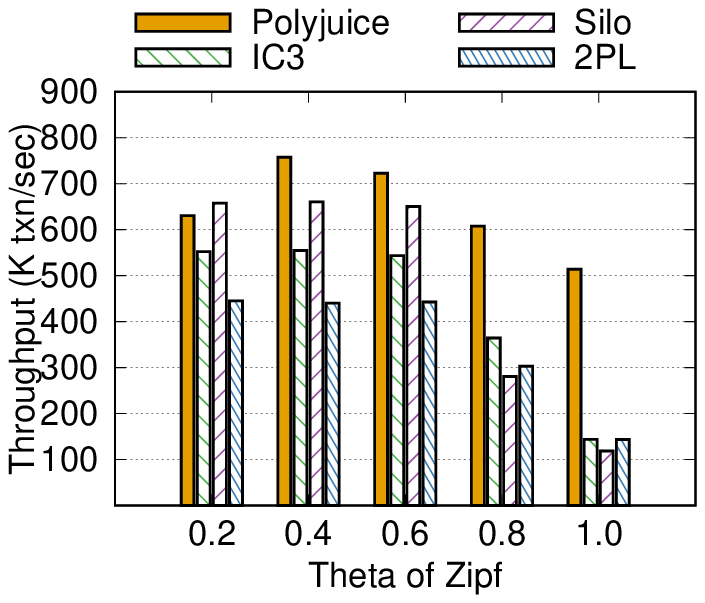}
\caption{Micro-benchmark with 10 tx types.}
\label{fig:ten-type}
\end{minipage}
\centering
\begin{minipage}[t]{0.22\textwidth}
\centering
\includegraphics[width=\textwidth]{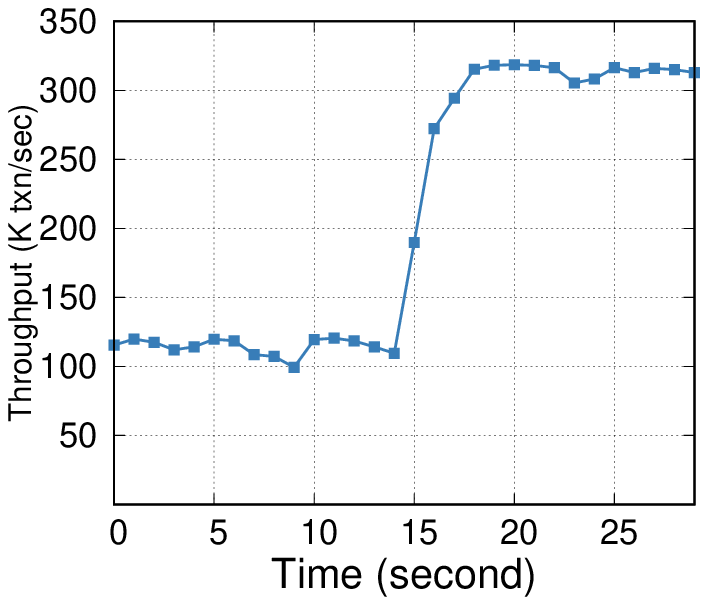}
\caption{Throughput during policy switch.}
\label{fig:switch}
\end{minipage}
\end{figure}

\subsection{Bigger benchmarks} 
We use two bigger benchmarks to check if \sys can learn a \cc policy in a much larger search space. The first benchmark includes 
three read-write transactions from TPC-E, TRADE\_ORDER, TRADE\_UPDATE and MARKET\_FEED. 
Compared with the state space of TPC-C (total 26 states), 
this benchmark is much more complex (total 65 states). 
The second benchmark is a micro-benchmark with ten types of transactions each with 8 accesses performing random updates (total 80 states). For each type of transaction, 
the last operation updates records in a unique table to distinguish it from other types. 
We build this benchmark because the action space grows exponentially 
with increasing transaction types. 

\begin{figure*}
\begin{minipage}{1\textwidth}
\centering
\subcaptionbox{Error rates for each day\label{fig:prediction-days}}{\includegraphics[width=.74\linewidth]{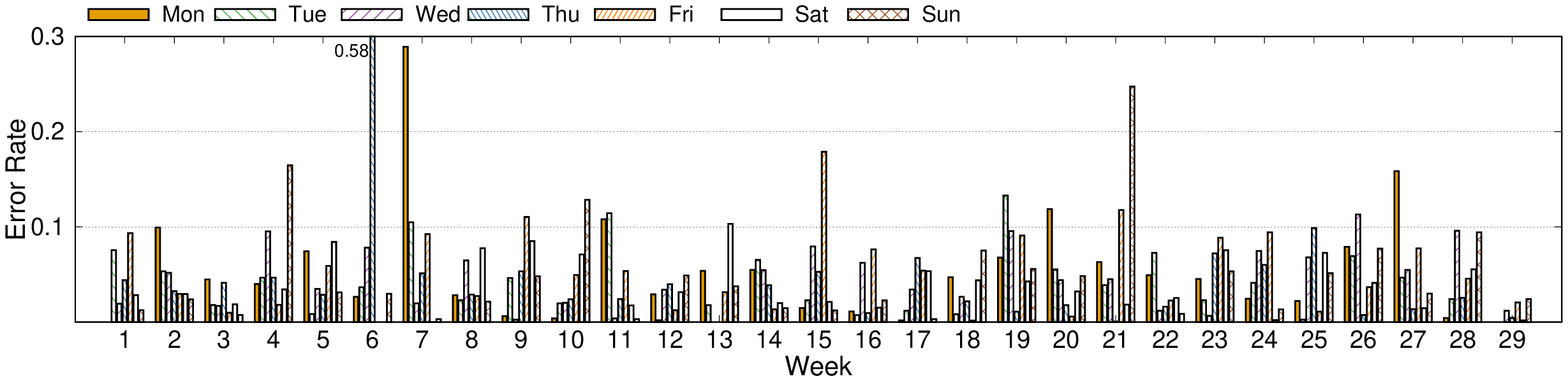}}
\subcaptionbox{CDF of error rates\label{fig:prediction-cdf}}{\includegraphics[width=.25\linewidth]{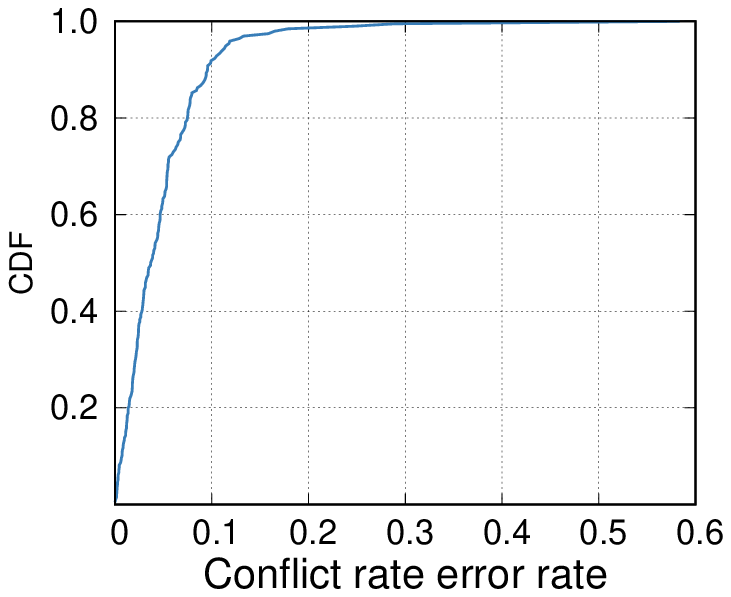}}
\caption{Error rates of conflict rate~\label{fig:predict}}
\label{fig:training}
\end{minipage}\hfill
\end{figure*}

\noindent \textbf{TPC-E.}
We vary the contention in TPC-E by controlling the updates on SECURITY table. 
Specifically, all updates follow the Zipf distribution and we vary the 
$\theta$ of Zipf from 0.0 to 4.0 to increase the contention. 
 We didn't evaluate Tebaldi as it doesn't provide a 
 manual grouping strategy for TPC-E. Similarly, we didn't evaluate CormCC as it is unclear how to partition the data for TPC-E.

As shown in \Cref{fig:tpce-performance}, the throughput of \sys is 42\%, 49\% and 55\% higher than other algorithms when contention is high ($\theta = 2, 3, 4$). Unlike TPC-C, in this experiment, the improvement of \sys is mainly attributed to the learned backoff. Specifically, \sys learns a different backoff mechanism from Silo's design. We find out that in Silo, the frequent aborts of TRADE\_ORDER result in a large backoff under high contention and the system spends a lot of time waiting before retry. In \sys, for TRADE\_ORDER transaction, it wouldn't increase the backoff even though the transaction is aborted. Although the abort rate remains high compared with Silo, the overall throughput is higher. \Cref{fig:tpce-scalability} shows the scalability of \sys under TPC-E with $\theta = 3$. \sys's performance can scale to 18.5$\times$ with 48 threads over that with a single thread, which is higher than IC3 (12.3$\times$) and 2PL (16.6$\times$). Silo (9.4$\times$) does not scale due to the frequent transaction aborts.

\noindent \textbf{Microbenchmark with 10 Types of Transactions.}
For this benchmark, we change the access distribution of the first operation to 
vary the contention level. Specifically, we change the $\theta$ of Zipf from 0.2 to 1.0 in the range of 4K. Other operations randomly update the records in the range of 10M, which results in little contention.
\Cref{fig:ten-type} shows the result, \sys's throughput is at least 66\% higher than other concurrency control mechanisms under high contention scenarios. This is because the learned policy pipelines the operations on some of the high-contention records while optimizing the waits for low-contention records.

\subsection{Training}
\label{subsec:training}

We have also implemented policy-gradient based RL training for the same workload. We initialize RL with an IC3-like policy to improve its training under this high contention workload. The initialization sets the parameters corresponding to IC3 actions with a high probability (in our case, 80\%).
The comparison result is shown in \Cref{fig:training-RL} for TPC-C with 1 warehouse and 48 threads. The RL agent converges after around 100 iterations, but the throughput of the learned policy is only 178K TPS. In contrast, \sys can learn a 309K TPS policy in 100 iterations. Our training runs on a single machine for now; each iteration takes $~$80 seconds, most of which are spent on evaluating policy performance.

\subsection{Coping with real-world workloads}
\label{subsec:dynamic-workload}

\subsubsection{Trace analysis}
\label{trace-analysis}
\textbf{The trace.} Our analysis is based on the trace of a real-world e-commerce website, downloaded from Kaggle~\cite{Kaggle}. 
The trace includes a log of requests sent to the web server, including the request time and several parameters. There are three types of requests: \texttt{VIEW}, for when a user views a product; \texttt{CART}, for when a user adds a product to the shopping cart; and \texttt{PURCHASE}, for when a user purchases a product. As \texttt{VIEW} corresponds to a read-only request,  we only include the two types of read-write requests \texttt{CART} and \texttt{PURCHASE} in our analysis.

\vspace{0.1in}
\noindent \textbf{Workload predictability.} For this analysis, we extract all the logged requests from Oct. 7th 2019 to Apr. 26th 2020 (29 weeks). After removing 6 invalid days, there are 197 days in total. We only consider the peak-hour workload for each day, since there is no need to optimize settings when the database is under-utilized and its commit throughput is limited by the incoming request rate instead of the \cc performance.

\begin{figure*}[t]
\begin{subfigure}[t]{0.99\textwidth}
\centering
\includegraphics[scale=0.66]{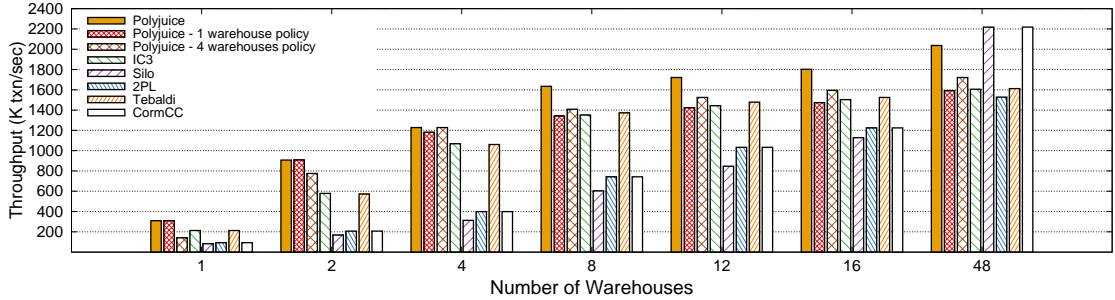}
\caption{TPC-C, 48 threads.}
\label{fig:sub-optimal-warehouse}
\end{subfigure}
\hfill
\begin{subfigure}[t]{0.99\textwidth}
\centering
\includegraphics[scale=0.66]{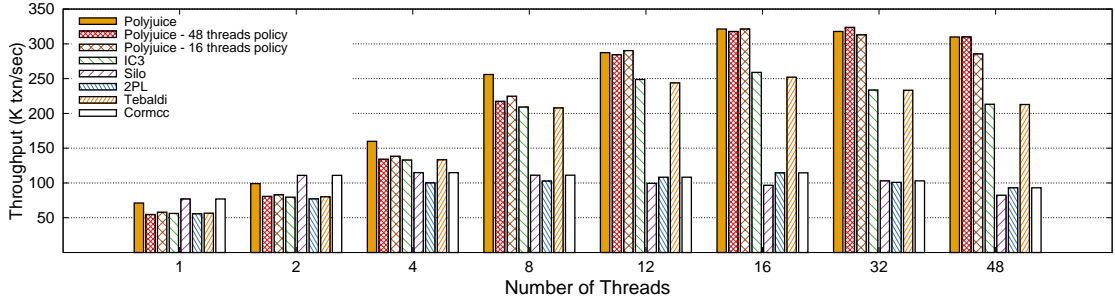}
\caption{TPC-C, 1 warehouse.}
\label{fig:sub-optimal-thread}
\end{subfigure}
\caption{Throughput under different workloads}
\end{figure*}

As proposed in \S~\ref{subsec:deployment}, we predict tomorrow's peak workload characteristic to be the same as today's peak. How accurate is such a prediction?  For our analysis, we characterize a workload by its contention level, which has the most effect on the learned policy. However, since the trace does not contain information on how long each request executes, we approximate the likelihood of contention by considering two requests to be in conflict with each other if they are sent by different users but operate on the same product id during some time window. We define $\mathit{conflict\_rate} = \mathit{conflict\_requests}/\mathit{total\_requests}$ within $n$ minutes. In our analysis, we set $n = 5$ and split an hour into 12 intervals. We use the mean of the 12 conflict rates to represent the contention in this hour and pick the hour with the most requests as the peak workload in a day. We note that $\mathit{conflict\_rate}$ is heavily influenced by the request rate; the bigger the request rate, the higher the measured conflict rate. 

\Cref{fig:predict} shows the error when predicting tomorrow's peak hour contention level using today's peak hour statistics. The error rate is calculated as $\mathit{error\_rate} = abs((\mathit{tomorrow} - \mathit{today})/\mathit{today})$. The smaller the $\mathit{error\_rate}$ is, the closer the next day's peak workload contention matches that of today. \Cref{fig:prediction-days} shows the error rate of the conflict rates for all 196 days (except for the first day), and \Cref{fig:prediction-cdf} shows the CDF of the error rates distribution. We can see that, there are only 3 days when the error rate of prediction is larger than 20\%. After manually checking these 3 days, we find out that they are due to a significantly higher or lower request rate, which affects the conflict rate.

We also analyze how frequently one needs to retrain. As suggested in \S~\ref{subsec:deployment}, we assume retraining is deferred until the predicted conflict rate differs from the one used for training the current policy by 15\%. For the trace analyzed, we only need to retrain 15 times to cover a period of 196 days. 

\subsubsection{Cost of policy switching}
\label{switching-cost}
We evaluate the cost of switching the policy in terms of: 1) how long it takes to fully switch the policy 2) whether commit throughput is affected by policy switching. The result is shown in ~\Cref{fig:switch}. We run the TPC-C 1 warehouse workload with 48 threads, and plot the throughput for each second. At the beginning, we run the workload with the OCC policy. Starting in the 15th second, we switch the policy to the one optimized for 1 warehouse. The result shows that it takes about 3 seconds to fully switch to a new policy, and switching does not negatively impact performance.  In fact, because we are switching to a better policy, the performance quickly improves during switching.

\subsubsection{Running a policy trained on a different workload}
\label{different-workloads}
We also study what happens if the workload optimized by the policy differs from the one actually being executed.  For these TPC-C experiments, we use fixed learned policies and measure their performance under various workloads that are different from those used in training. 

In the first set of experiments, we 
use two fixed policies, which are trained using 48 threads on 1 warehouse or 4 warehouses.
 \Cref{fig:sub-optimal-warehouse} shows 
the performance of fixed policies as we vary the number of warehouses, compared to existing algorithms and \sys when it is always trained on the correct workload. If the evaluation workload is  different from the workload used 
for training, the fixed policies can be  sub-optimal. For example, the performance 
of \sys (1-warehouse) is 71\% of Silo under 48 warehouses. However, the performance differences between fixed and optimal policies are small when the evaluation workload is not too far off from the training workload. 

In the second set of experiments, we use fixed policies trained on 1 warehouse using 48 or 16 threads.  \Cref{fig:sub-optimal-thread} shows the performance of fixed policies as we vary the number of threads.  The results are similar, in that a trained policy is fairly robust to  training and evaluation workload mismatch. 

\section{Related Work}
\label{sec:related}

\noindent \textbf{Concurrency control.} 
We can categorize recent \cc works according to their 
design choices. 
1) Scheduling based CC: IC3~\cite{wang2016scaling}, Callas~\cite{xie2015high},
DPR~\cite{mu2019deferred} and RoCoCo~\cite{mu2014extracting} allow ongoing transactions to expose their writes 
and track dependencies at runtime, then schedule the read/write operations according to the tracked dependencies. 
Ding et al.~\cite{ding2018improving} schedules read operation after conflicting transaction's 
commits to avoid aborts for OCC protocol. 
2) Deterministic databases:
Granola~\cite{cowling2012granola}, Deterministic 
CC~\cite{thomson2012calvin, ren2019slog, faleiro2014lazy, ren2016design} and Eris~\cite{li2017eris} schedule a transaction's 
execution according to a predetermined order. PWV~\cite{faleiro2017high} adds early write visibility to the 
deterministic \cc to further improve the performance. 
3) Changing the validation algorithm to avoid unnecessary aborts: 
TicToc~\cite{yu2016tictoc} avoids unnecessary aborts by using logical timestamps for validation. BCC~\cite{yuan2016bcc} changes the validation phase by detecting a special pattern. 
4) Partially rolling back to reduce the abort cost ~\cite{wu2016transaction}.

In addition, there are a number of works applying MVCC into their systems. 
Bohm~\cite{Faleiro14vldb} combines the MVCC with deterministic CC to achieve non-blocking 
operations. Cicada~\cite{lim2017cicada} uses logical timestamps with MVCC to increase the possibility of constructing safe interleavings.  Obladi~\cite{crooks2018obladi} integrates MVCC on top of ORAM to provide security along with high performance.

All above CC protocols leverage a fixed set of design choices. Compared to them, \sys is able to adapt the 
design choices according to the characteristics of a given workload.
Some work~\cite{szekeres2020meerkat, zhang2018building, mu2016consolidating} focus on distributed databases, which must do replication in addition to concurrency control. They propose new algorithms to handle inconsistent orderings in both concurrency control and replication.

\noindent \textbf{Hybrid concurrency control.}
There are existing works that combine multiple concurrency control mechanisms for better performance. 
MOCC~\cite{wang2016mostly} develops a specific algorithm to combine OCC and 2PL for high-contention workloads. Sundial~\cite{sundial:vldb18} proposes a new hybrid CC algorithm based on 2PL 
and OCC with logical timestamps. CormCC~\cite{tang2018toward} proposes a more general hybrid method by 
formalizing all CC into four phases. Each operation can use any CC's policy as long as all CCs perform 
each phase according to the same order. Tebaldi~\cite{su2017bringing} groups transactions and assigns different \cc protocols to each group.
However, existing algorithms are either specific for combining
OCC or 2PL, or need programmers to provide heuristics to choose the execution policy for each operation. 
Compared to them, \sys is able to automatically adapt the policy for each operation according to 
the workload.

\noindent \textbf{Learned systems.}
Many system optimizations can be done by machine learning models trained from historical data.
In the area of databases, examples include cardinality estimation
~\cite{lakshmi1998selectivity, kipf2018learned, wu2019towards, park2018quicksel}, join order
planning~\cite{krishnan2018learning, marcus2018deep, ortiz2018learning} and
configuration tuning~\cite{van2017automatic}. Besides databases,
works have been done to improve buffer management systems
\cite{chendeepbm}, sorting algorithms \cite{zhao2018n}, memory page prefetching
\cite{hashemi2018learning, zeng2017long} and memory control \cite{ipek2008self}, task scheduling \cite{kraska2019sagedb}, CPU scheduling \cite{sheng2019scheduling}, locking priority \cite{eastep2010smartlocks} and cache replacement ~\cite{song2020learning}. 
Although these works try to leverage machine learning to make systems self-aware, but none of them targets on the concurrency control. Thus, they have different model design from \sys.

\section{Discussion}
\label{sec:limit}

As a first attempt on learnable \cc, 
\sys has limitations, some of which we hope to address in the future.

\noindent \textbf{Not suitable for rapidly changing workloads.} In our experience, training takes on the order of several hundred seconds. Thus, \sys is not suitable for scenarios in which workload changes quicker than every few minutes.

\noindent \textbf{Inaccurate workload emulation.}  Training reissues executed transactions with their logged inputs. However, since transaction interleavings during training differ from that of the original execution, a transaction's outputs also differ.  \sys works only if such emulation inaccuracies do not significantly affect the workload access pattern.

\noindent \textbf{Large state space.} \sys represents $a^s$ potential policies in a table format, 
where $s$ is the number of different states and $a$ is the number of different actions per state.
As the number of transactions and the number of accesses in each transaction increase in the workload, 
both $s$ and $a$ increase.  The resulting much enlarged search space will make training via EA less effective.  
One potential solution is to follow the breakthrough of deep reinforcement learning, and 
use a function approximator like a deep neural network to approximate the policy table with parameters 
far fewer than the number of table cells. It is a well-known challenge to make deep RL work 
effectively.

\noindent\textbf{More expressive policy space.} There are several interesting directions to expand the policy space, such as 
supporting multi-version databases, explicit 
CPU scheduling of execution, fine-grained instead of binary contention levels.

\noindent\textbf{Weaker and mixed isolation levels.} \sys currently only guarantees serializability. Some applications can work with weaker or mixed isolation levels~\cite{weakisolation:popl18, xie2014salt, mehdi2017can, lloyd2011don, crooks2016tardis}.  It is an interesting extension to generalize to these scenarios.
\section*{Acknowledgements}

Chien-chin Huang and Minjie Wang contributed valuable ideas in the early stage of this project. We thank the anonymous reviewers for the valuable comments. We are especially grateful to our shepherd, Deniz Alt{\i}nb{\"u}ken, for helping improve the paper's presentation. 
Jiachen Wang, Huan Wang, Zhaoguo Wang and Haibo Chen were supported by National Key Research and Development Program of China (No. 2020AAA0108500), National Natural Science Foundation of China (No. 61902242), and the HighTech Support Program from Shanghai Committee of Science and Technology (No. 20ZR1428100). Ding Ding, Conrad Christensen, and Jinyang Li were supported by NSF grant 1816717, and a gift from NVIDIA and AMD. Zhaoguo Wang (zhaoguowang@sjtu.edu.cn) and Jinyang Li (jinyang@cs.nyu.edu) are the corresponding authors.


\bibliographystyle{plain}
\bibliography{chamcc-paper}

\ifx \arxiv\mode
\clearpage
\appendix
\begin{algorithm}[H]
\caption{Transaction execution in \sys}
    \label{alg:code}
 \begin{algorithmic}[1]
 
\Function{Put}{$k$, $v$, $T$, $tx$-$type$, $acc$-$id$}
\State {// $k$: key, $v$: value, $T$: transaction object}
\State {// $tx$-$type$: transaction type, $acc$-$id$: access-id}
  \State $r$ $\leftarrow$ db.Lookup($k$)
  \State action $\leftarrow$ policy.Lookup($tx$-$type$, $acc$-$id$, ACCESS)
  \State WaitUntil(action.waits)
  \State $T$.buffer.append($k$,$v$, WRITE)
  \State $T$.wset.append($k$,$v$)
  \If{action.write\_visible == PUBLIC}
   \If {Early\_Validate($T$.buffer, $acc$-$id$) fails}
  \State rollback $T$.wset, $T$.rset
  \State  $T$.buffer $\leftarrow$ $\{\}$
  \State  goto last point of successful validation
   \Else { AppendToAccessList($T$.buffer)}
  \EndIf
  \State  $T$.buffer $\leftarrow$ $\{\}$
  \EndIf
\EndFunction

\Function{Get}{$k$, $T$, $tx$-$type$, $acc$-$id$}
\State $r$ $\leftarrow$ db.Lookup($k$)
 \State action $\leftarrow$ policy.Lookup($tx$-$type$, $acc$-$id$, ACCESS)
 \State WaitUntil(action.waits)
  \If{action.read\_version == DIRTY\_READ}
  \State $v$ $\leftarrow$ FindLastWrite($r$.acc\_list)
  \Else { $v$ $\leftarrow$ $r$.data}
  \EndIf
  \State $T$.buffer.append($k$, $v$, READ)
  \State $T$.rset.append($k$,$v$)
  \If {action.early\_validate}
  \If {Early\_Validate($T$.buffer, $acc$-$id$) fails}
  \State rollback $T$.wset, $T$.rset
  \State  $T$.buffer $\leftarrow$ $\{\}$
  \State  goto last point of successful validation
   \Else { AppendToAccessList($T$.buffer)}
  \EndIf
  \State  $T$.buffer $\leftarrow$ $\{\}$
  \EndIf
\EndFunction
  
\Function{Early\_Validate}{$T$.buffer, $acc$-$id$}
 \State action $\leftarrow$ policy.Lookup($tx$-$type$, $acc$-$id$, VALID)
 \State WaitUntil($T$.deps)
 \State Validate($T$.rset, $T$.wset)
\EndFunction

\Function{Commit}{$T$, $tx$-$type$}
 \State {// Add all dirty reads in $T$.buffer to $T$.deps.}
 \State WaitUntil($T$.deps)
\If {Validate($T$.rset, $T$.wset) fails}
\State retry from beginning
\Else{ atomically write $T$.wset to db}
\EndIf
\EndFunction
\end{algorithmic}
\end{algorithm}

\section{Proof}
\label{sec:proof}

We give a sketch of the correctness argument. \\\\
1) Polyjuice never commits a transaction that has read data from some aborted transactions.

This is because if a read operation passes validation, the read version must be committed (Lemma-1).\\\\
2) All committed transactions are serializable. 

We use proof-by-contradiction to show there cannot be cycles in any serialization graph (Lemma-3).

\begin{lemma}
\textbf{commit read validation:} for transaction $T_r$, if $T_r$ reads the value of R written by $T_w$ with version $v_w$ (both committed or uncommitted), then $T_r$ can successfully pass the validation on R if only if $T_w$ finally commits version $v_w$ of R.
\end{lemma}
\textbf{Proof} Assume $v_w$ hasn’t been successfully committed (including Subcase-1: $T_w$ has aborted itself, Subcase-2: $T_w$ has been successfully committed, but $T_w$ commits another version $v_w’$ (different from $v_w$) on R.). There doesn’t exist another version $v’$ of R which shares the same version-id as $v_w$ (Lemma-2), and thus $T_r$ cannot pass the validation on R.

$\square$

\begin{lemma}
\textbf{unique version-id:} for any record, there doesn’t exist two different versions(including both committed and uncommitted version) $v_1$ != $v_2$, such that $v_1$’s version-id == $v_2$’s version-id.
\end{lemma}
\textbf{Proof} Assume $v_i$ and $v_j$ are version-ids of two different versions to some record, created by $T_i$ and $T_j$. Specifically, $v_i$ includes the txn-id of $T_i$ and the seqno if $T_i$ publics this version before commit, ditto for $v_j$.
\begin{outline}
\1 Case-1 $T_i$ != $T_j$, we never assign a txn-id twice.
\1 Case-2 $T_i$ == $T_j$, since a txn never commits twice, there must be a version published before $T_i$’s commit. For each version published before commit, we assign a unique seqno. 
\end{outline}
$\square$

~\\[3pt]
\textbf{For the proof of serializability,  we also use the following definitions:}

\textbf{Committed read/write:} the read/write value associated with a committed txn.

\textbf{Serialization graph:} the graph consists of only committed transactions.  The edges in the graph have 3 types, based on conflicts between committed reads/writes.

$T_i$ $\xrightarrow{ww}$ $T_j$: $T_j$’s committed write overwrites $T_i$’s committed write to the same record in the data store. 

$T_i \xrightarrow{wr} T_j$: $T_j$’s committed read is the same as $T_i$’s committed write to the same record in the data store.

$T_i \xrightarrow{rw} T_j$: $T_j$’s committed write overwrites the version which is $T_i$’s committed read to the same record in the data store.

\textbf{Serializability:} there is no cycle in the serialization graph.\\\\

\textbf{Proof sketch:} Suppose there exists some violation of serializability, thus there exists a cycle in some serialization graph.  Suppose the cycle is $T_1$ $\rightarrow$ $T_2$ $\rightarrow$ … $\rightarrow$ $T_n$ $\rightarrow$ $T_1$. We use a lemma (proven later) that if $T_i$ $\rightarrow$ $T_j$, then $T_i$ should acquire all locks (end-of-lock-stage) in the commit phase before $T_j$  acquire all locks (end-of-lock-stage). Let’s use < to indicate happens-before relationship. Thus, $T_1$’s end-of-lock-stage < $T_2$’s end-of-lock-stage < … < $T_1$’s end-of-lock-stage, which forms a contradiction.

\begin{lemma}
Given an edge $T_i$ $\rightarrow$ $T_j$ in the serialization graph, then $T_i$’s end-of-lock-stage < $T_j$’s end-of-lock-stage, a.k.a. $T_i$ acquires all locks before $T_j$ acquires all locks.
\end{lemma}

We prove this case by case, according to 3 edge types: $\xrightarrow{ww}$, $\xrightarrow{wr}$ or $\xrightarrow{rw}$. For concreteness, let’s assume the conflict edge is associated with record R.

\begin{outline}
\1 Case-1 is for $T_i$ $\xrightarrow{ww}$ $T_j$.  This implies “$T_i$ acquires R’s lock < $T_i$ writes to R < $T_i$ releases R’s lock < $T_j$ acquires R’s lock”, and thus we can have “$T_i$’s end-of-lock-stage < $T_j$’s end-of-lock-stage”.
\1 Case-2 is for $\xrightarrow{wr}$ $T_j$. 
    \2 Subcase-2-1 is for $T_j$ reads the clean (committed) value of R, which is committed by $T_i$. This implies $T_i$’s acquires R’s lock < $T_i$’s write  < $T_j$’s read, which happens before $T_j$ starts its commit phase, and thus $T_i$’s end-of-lock-stage < $T_j$’s end-of-lock-stage.
    \2 Subcase-2-2 is for $T_j$ reads the dirty (uncommitted) value of R, which is finally committed by $T_i$. Since $T_j$ will wait for all the direct dependent transactions to commit/abort before it enters the commit phase, $T_j$ needs to wait for $T_i$ to finish commit before $T_j$ starts its commit phase, and thus $T_i$’s end-of-lock-stage < $T_j$’s end-of-lock-stage.
\1 Case-3 is for $T_i$ $\xrightarrow{rw}$ $T_j$. Let’s assume  $T_i$ read the version installed by $T_c$.
To prove “$T_i$’s end-of-lock-stage < $T_j$’s end-of-lock-stage”, we first prove that “$T_i$’s version check on R < $T_j$ acquires R’s lock”. 
    \2 Subcase-3-1 is for $T_j$ releases R’s lock < $T_i$’s version check on R, which means $T_j$ commits its writes to R before $T_i$ validates R.  According to the definition, $T_i$ $\xrightarrow{rw}$ $T_j$ implies that $T_j$ overwrites $T_c$’s committed write, which is read by $T_i$ (Lemma-1). Thus, we have  “$T_c$’s write < $T_j$’s write < $T_i$’s version-check”. For this order, $T_i$ will fail to validate R, this is because $T_j$ installs a version id different from $T_c$ (Lemma-2) before $T_i$’s validation. 
    \2 Subcase-3-2 is for $T_i$’s version check on R < $T_j$ releases R’s lock. This includes the following two cases: 
    \begin{itemize}
        \item First, “$T_j$ acquires R’s lock < $T_i$’s version check on R”.
        \item Second, “$T_i$’s version check on R < $T_j$ acquires R’s lock”.
    \end{itemize}
    Then we prove that the first case is impossible: if “$T_j$ acquires R’s lock < $T_i$’s version check on R”, then we have $T_j$ acquires R’s lock < $T_i$’s version check < $T_j$ releases R’s lock. For this order, $T_i$ will fail to validate R, this is because $T_i$ would abort itself upon encountering $T_j$’s lock on R.
\end{outline}
Therefore, we prove that $T_i$’s version check on R < $T_j$ acquires R’s lock, which implies $T_i$’s end-of-lock-stage < $T_j$’s end-of-lock-stage.

$\square$
\fi

\ifx \camera\mode
\clearpage
\input{osdi21_ae_appendix_template}
\fi
\end{document}